\documentclass[conference]{IEEEtran}
\IEEEoverridecommandlockouts
\usepackage{cite}
\usepackage{amsmath,amssymb,amsfonts}
\usepackage{algorithmic}
\usepackage{textcomp}
\usepackage{xcolor}

\usepackage{graphicx}
\usepackage{multirow}
\usepackage{siunitx}
\usepackage[vlined,linesnumbered,ruled]{algorithm2e}
\makeatletter
\renewcommand{\@algocf@capt@plain}{above}
\makeatother
\usepackage{subcaption}
\usepackage[most]{tcolorbox}
\usepackage[export]{adjustbox}
\usepackage{array}
\usepackage{rotfloat}
\usepackage[T1]{fontenc}
\usepackage[latin9]{inputenc}
\usepackage{amssymb}
\usepackage{listings}
\usepackage{xcolor}

\definecolor{codegreen}{rgb}{0,0.6,0}
\definecolor{codegray}{rgb}{0.5,0.5,0.5}
\definecolor{codepurple}{rgb}{0.58,0,0.82}
\definecolor{backcolour}{rgb}{0.95,0.95,0.92}
 
\lstdefinestyle{code_style}{
    backgroundcolor=\color{backcolour},   
    commentstyle=\color{codegreen},
    keywordstyle=\color{magenta},
    numberstyle=\tiny\color{codegray},
    stringstyle=\color{codepurple},
    basicstyle=\ttfamily\tiny,
    breakatwhitespace=false,         
    breaklines=true,                 
    captionpos=b,                    
    keepspaces=true,                 
    numbers=left,                    
    numbersep=5pt,                  
    showspaces=false,                
    showstringspaces=false,
    showtabs=false,                  
    tabsize=2
}
 
\def\BibTeX{{\rm B\kern-.05em{\sc i\kern-.025em b}\kern-.08em
    T\kern-.1667em\lower.7ex\hbox{E}\kern-.125emX}}
    
\begin{document}

\title{An Automated Testing Framework For Smart TV apps Based on Model Separation
{\footnotesize }
}

\author{\IEEEauthorblockN{Bestoun S. Ahmed}
\IEEEauthorblockA{\textit{Dept. Mathematics \& Computer Scie.} \\
\textit{Karlstad University}\\
Karlstad, Sweden \\
bestoun@kau.se}
\and
\IEEEauthorblockN{Angelo Gargantini}
\IEEEauthorblockA{\textit{Dept. Manage. Info. \& Production Eng.} \\
\textit{University of Bergamo}\\
Bergamo, Italy \\
angelo.gargantini@unibg.it}
\and
\IEEEauthorblockN{Miroslav Bures}
\IEEEauthorblockA{\textit{Dept. Computer Scie.} \\
\textit{Czech Technical University}\\
Prague, Czech Republic \\
buresm3@fel.cvut.cz}

}

\maketitle

\begin{abstract}
Smart TV  application (app) is a new technological software app that can deal with smart TV devices to add more functionality and features. Despite its importance nowadays, far too little attention has been paid to present a systematic approach to test this kind of app so far. In this paper, we present a systematic model-based testing approach for smart TV app. We used our new notion of model separation to use sub-models based on the user preference instead of the exhaustive testing to generate the test cases. Based on the constructed model, we generated a set of test cases to assess the selected paths to the chosen destination in the app. We also defined new mutation operators for smart TV app to assess our testing approach. The evaluation results showed that our approach can generate more comprehensive models of smart TV apps with less time as compared to manual exploratory testing. The results also showed that our approach can generate effective test cases in term of fault detection.

\end{abstract}


\section{Introduction}

Smart TV application (app) is a software designed to run on smart TV devices. In addition to the television service, smart TVs add many functionalities and the ability to connect to the Internet. Nowadays, smart TVs are going to dominate the television market, and the number of connected TVs is growing exponentially. This growth is accompanied by an increase in the number of consumers and the use of smart TV apps that drive these devices.

In general, software apps testing of smart devices can be considered as an extension and evolution of testing practices from the traditional user interfaces (UI) like the graphical user interface (GUI) of desktop and web apps. The testing practices for these GUIs have been studied extensively in the last decade, and as a result, many sophisticated methods, algorithms, and tools have been developed. Banerjee \textit{et al.} \cite{Banerjee:2013:GUI} studied more than 230 articles published between 1991-2013 in the area of GUI testing and Li \textit{et al.} \cite{Li:2014:TDW} surveyed the literature in two decades of web app testing. As for smart devices, Zein \textit{et al.} \cite{ZEIN2016334} have studied the mobile apps testing area through 79 articles published between 2005-2015. It is noticeable that in the three types of apps (i.e., desktop, web, and mobile apps), model-based testing is the common approach among the leading testing technologies in the literature. We also noticed a lot of overlap among the model-based testing approaches of GUI. For example, the MobiGUITAR strategy \cite{Amalfitano2015MobiGuitar} for systematic mobile app testing has been inspired by the GUITAR strategy \cite{Nguyen2014:Guitar} for desktop GUI testing. Here, the model is constructed using the user interaction to create the interaction events; then the model used to generate the test cases. The finite-state machine (FSM) and directed graphs were used widely in the literature as a model abstraction to represent the state and events on the GUI. Despite these common aspects of the testing, due to the different interaction mode, it is impossible to have a comprehensive strategy for all types of apps. In other words, it is not possible to use the same framework to test mobile, web, and desktop apps.


With the increasing importance and use of smart TV apps, there is an urgent need to have a testing framework for these apps. It is becoming increasingly difficult to ignore the testing of smart TV apps and assuming that they have the same testing procedure that we usually use for mobile apps. While there are many analogies between the two kinds of apps, still there is a great deployment effort estimated to migrate from mobile to smart TV app testing due to the different mode of the user interaction. Although many strategies can be found for mobile apps testing, so far, there has been no discussion about smart TV apps. In fact, far too little attention has been paid to test this kind of app and so far, we cannot find a systematic framework to test it. An essential contribution of this paper is to identify those analogies and differences between the testing procedure for these two types of apps. Recently, we have addressed the key ingredients and challenges for smart TV apps testing \cite{BestounSmartTVConfPaper}. We also proposed some preliminary solutions for those problems.

In this paper, we aim to establish the research foundation of smart TV app testing. To achieve this aim, we developed a black-box model-based testing framework called EvoCreeper to test smart TV apps. Our approach is to explore the GUI of the smart TV app by examining each element and monitoring its reaction. The testing framework takes the TV app to automatically create a complete model by examining the GUI widgets and the transitions among them; then, it generates the test cases based on this model. The framework also supports test case execution and verification. While we developed new algorithms for this testing framework, still, we deployed many algorithmic concepts (wholly or partially) from the earlier methods and practices used for GUI, web, and mobile apps.

The framework constructs a model for a given smart TV app cumulatively by exercising the user interface and extracting the runtime information using the remote control device interaction with the app. An important initial stage of the proposed testing framework is the black-box reverse engineering stage that is used to generate a comprehensive model for the app-under-test. The generated model can also be used for several verification and validation activities during the development process.


Another significant contribution of this paper is the idea of test case generation based on sub-model separation. Using the FSM and directed graphs as a model to represent the widgets and events on the GUI leads to a graph with many nodes and edges. The nodes are the widgets on the GUI and the edges are the transition among them. Due to the different nature of smart TV apps, the number of edges and nodes will be much higher than other apps. This will lead to a combinatorial explosion problem in the model in which we cannot generate all the test cases. One possible solution towards the optimization of the test cases is to use search-based algorithms to cover all the nodes and edges by the smallest set of test cases. However, in many instances, these algorithms are computationally expensive for such a complex model. Besides, search-based algorithms are more suitable for comprehensive testing of the app. Also, for this kind of model, where the sequence of the states is crucial, using search-based algorithms may not lead to the correct sequence of events on the smart TV app when it comes to executing the test cases due to the generation of unfeasible transition. As in the case of desktop GUI testing, either the generated test suite must be repaired or a constraint solver must be used. Both solutions are computationally expensive. In this paper, we introduce a new model-based prioritization method to generate customized effective test cases by extracting the user preferred sub-model from the exhaustive model (we call it Mega-model). We have demonstrated the preliminary design of this model constructor in our recently published study \cite{EvoCreeperRef}. Using this method, more effective and efficient test cases can be generated by giving the authority to the tester to choose the path that he wants to test.

\section{Motivation and Related Work}\label{MotivationRelatedWork}

Model-based concepts have been used frequently in the literature to test the GUI of different software apps to assess their functionality. Most of the studies simulated the user interaction behavior with the GUI to construct the model. Here, reverse engineering has been used as an alternative to the classical capture/replay tools \cite{Hicinbothom1993} to build a model. The model is then used to generate different test cases that represent sequences of execution to be run on the GUI for verification. Typically, the model is presented as nodes and edges. The nodes represent all the building blocks (i.e., widgets) of the GUI, (e.g., buttons, menus) and the edges represent the transitions among the nodes and the  GUI's execution behavior (e.g., click, touch). Using this approach, Memon \textit{et al.} \cite{MemonReverse2003,Nguyen2014:Guitar} developed GUITAR, a reverse engineering framework to model the GUI of the desktop software apps. GUITAR generates an event-flow graph (EFG) by capturing the widgets on the GUI. An algorithm extracts all the window's building blocks of the GUI, their values and properties. The capturing algorithm also captures different widgets that leads to obsolete test cases. For example, when the label widget is not clickable and presenting it in a test case leads to non-executable test. Here, an algorithm has been used in GUITAR to repair those obsolete test cases. This approach has undergone many development stages in different studies. Aho et al. \cite{AHO201449} presents an extensive survey for these studies and the techniques.

In line with this approach, Mesbah \textit{et al.} \cite{Mesbah:2012} suggested CRAWLJAX, a web app's reverse engineering method to generate models that relied on the dynamic crawling on the app to detect the clickable widgets. In addition to the dynamic crawling on the app, CRAWLJAX also relies on the code scanning of the web app to fine-tune the model. The model then used to generate the necessary executable test cases. This approach also motivated other researchers to use it for mobile app model-based testing. Joorabchi and Mesbah \cite{Joorabchi2012IOS} developed a reverse engineering method to construct the model for the GUI of mobile apps.

With the different app types and different software technologies, in most cases, the same approach has been applied in the literature. However, it is the technology, app type, and user interaction that makes the difference when applying the approach. Within each app type, there is a set of difficulties that alter this model-based approach to be different. Due to these differences, especially in GUI technology, it is challenging to develop a generic testing framework that can generate models and test cases for different kind of GUI-based apps. 

The smart TV app is another new kind of smart device app that relies on a fully interactive GUI. Smart TV devices are becoming more popular these days as there are a lot of advancements in the device and app development. Nowadays, smart TV devices are actively involved in other technologies like smart home, Internet of things (IoT), and controlling other consumer electronics. For example, smart security camera system \cite{Erkan2015}, home sleep care with video analysis using smart TV app  \cite{Fan2014}, controlling the smart home from smart TV \cite{Cabrer2006}, Smart Lighting Control \cite{Chun2013}, healthcare app of smart home \cite{Vavilov2014}, home appliances control through Smart TVs\cite{Kim2013,Kim2015}, and many other apps. It is strongly become difficult to ignore the testing of the smart TV app. Using model-based testing approach for smart TV app is relatively under-explored in the literature.

Although there are a lot of synergies between the mobile and web apps' testing, still, there are many expected difficulties when one tries to apply the same model-based approach due to the different user interaction with the GUI and the various technologies that the smart TV app build on. Mainly, using the remote control device to interact with the app will change the model. The widgets representation on the GUI and building technology behind that also will change the model of the app. Applying the same reverse engineering methods used for the web and mobile apps will lead to creating a useless model when it comes to executing the test cases due to the different user interaction. When modeling the web app, the model is prepared for the interaction of mouse and keyboard inputs. In the same way, the model for the mobile app is prepared for the interaction of the touch screen with the fingers for example. However, smart TV app relies on the remote control device which behaves differently. For example, the mouse or finger can move in different directions that lead to moving from any widget to another for an event. Whereas, the remote control device moves only in four directions (Left, Right, Up, and Down).

The user interaction behavior has a direct impact on the type of the model graph generated for testing in the literature. For example, Nguyen \textit{et al.} \cite{Nguyen2014:Guitar} constructed the EFG as a model for the desktop GUI testing, whereas, Amalfitano \textit{et al.} \cite{Amalfitano2015MobiGuitar} constructed the state machine graph as a model for mobile app testing. For smart TV apps, both the EFG and state machine graph models are not applicable. Here, each move from a node to another on the graph is a step. Hence, the weight of an edge between two nodes on the graph is the same. While this weight of the edge is not essential for the other apps, it is essential in the smart TV app. For example, while the user can move from an icon to another in the mobile app, there are restrictions of movement in the smart TV apps. Here, we assume the regular daily use of the apps. Nowadays, from a typical personal computer (PC) to smart TV, most of the devices are deployed with a touchscreen. However, still the normal daily user interaction with the PC app is the mouse device and with the TV is the remote control device. 

Recently, we have tried to study and summarize the key problems of testing smart TV apps \cite{BestounSmartTVConfPaper}. In our study, we have also tried to give different suggestions to solve those problems without implementing them. In another recently published study \cite{EvoCreeperRef}, we tried to construct a preliminary model smart TV apps, without using the model in any testing activities. A significant effort can be found by Cui \textit{et al.} \cite{Cui2017} to address the testing of the smart TV app. The study follows the previous approaches of model extraction by exhaustively explore the Android smart TV app to detect all the GUI widgets then optimize and filter them for the model construction. Here, a significant effort is needed for model optimization and reduction. The widgets extraction used a white-box crawler to scan the code of the app to construct a preliminary model, that contains many obsolete nodes. These obsolete nodes are originated from the detection of all the widgets in the crawling of the code regardless if the widgets are clickable or not. For example, the crawler detects a text label on the GUI as a node in the model while it is not active as it is not clickable. The study proposed an algorithm to reduce those obsolete nodes by deleting the inactive (or obsolete) nodes. Although promising, as mentioned previously, this approach has many drawbacks. When the app size become bigger, this process becomes time-consuming and there is a great possibility to stuck in the combinatorial explosion problem in the constructed model. Also, there is a need to formulate the model for the complex structure of different apps. The study did not show the detail of the testing process nor named the apps used for the case studies. It is not clear also how the test cases generated from the constructed model and what kind of assessment used to evaluate the effectiveness of the generated test cases.

In contrast to the Cui \textit{et al.} study \cite{Cui2017}, our approach uses a more effective method by only detecting the active widgets of the GUI, which in turn does not need to reduce the model. Hence, the constructed model by our approach is a more realistic model and there is no need for an addition algorithm to repair the generated test cases from it. Another contribution of our approach is that it will consider the actual movement of the remote controller device and does not rely on the code scanning of the app. Hence, in contrast to the study by Cui \textit{et al.}, our approach uses the black-box approach in which it will construct the model for the app, even its code is not available. Here, the GUI is explored using the remote control device while examining each widget on it and listen to the reaction for recording. When the widget element is clickable, the algorithm will consider it as a state in the model. The transition among the recorded states then represent the edges in the model.

\section{Smart TV App Technology and Its Testing Challenges}\label{SmartTVAppTestingChallenges}

As in the case of the other new smart apps nowadays, Software Development Kits (SDK) used to develop smart TV apps. From the development perspective, each smart TV platform used to have its SDK. Cross-platform frameworks like Joshfire\footnote{https://www.joshfire.com/}, Mautilus\footnote{https://www.mautilus.com/}, and Smart TV Alliance\footnote{http://www.smarttv-alliance.org/} were also an option to develop apps. However, the apps developed by them were able to work on a few commercial TV devices. Besides, some of those projects were shut down after a while. Tizen studio SDK nowadays used as an effective development framework that depends on the latest web technology and has a set of supporting tools for development. Tizen studio used to develop apps for the Tizen operating system for smart TV devices. To support cross-platform apps, Tizen used the most up-to-date web packages such as CSS, HTML5, JavaScript and W3C widget.

The user primarily interacts with the smart TV app through the remote control device. Although many smart TV devices are nowadays supporting the direct interaction with the touchscreen, still, the 10-feet (3m) distance is the standard distance of the user from the TV device \cite{Sabina2016,OperahDesignDevelope}. In fact, the remote control device is not as easily as used like fingers or mouse device for example. To this end, developers always consider this situation when designing the GUI of smart TV apps. For example, the developers are avoiding the use of text boxes as much as they can when designing the GUI of the smart TV app because entering text is complicated and not user-friendly. There are six essential and constant buttons on the device that any smart TV must have to deal with the GUI. Those buttons are four navigation buttons (Right, Left, Up and Down), Back to navigate back to the previous state on the GUI, and the OK to select an item on the GUI. There are other buttons on the remote devices for some functional reasons such as on/off buttons or a set of (0-9) numbers for channel jumping, but they are not affecting the navigation.

\begin{figure*}
	\centering
	\begin{subfigure}[b]{0.3\textwidth}
		\includegraphics[width=\textwidth]					{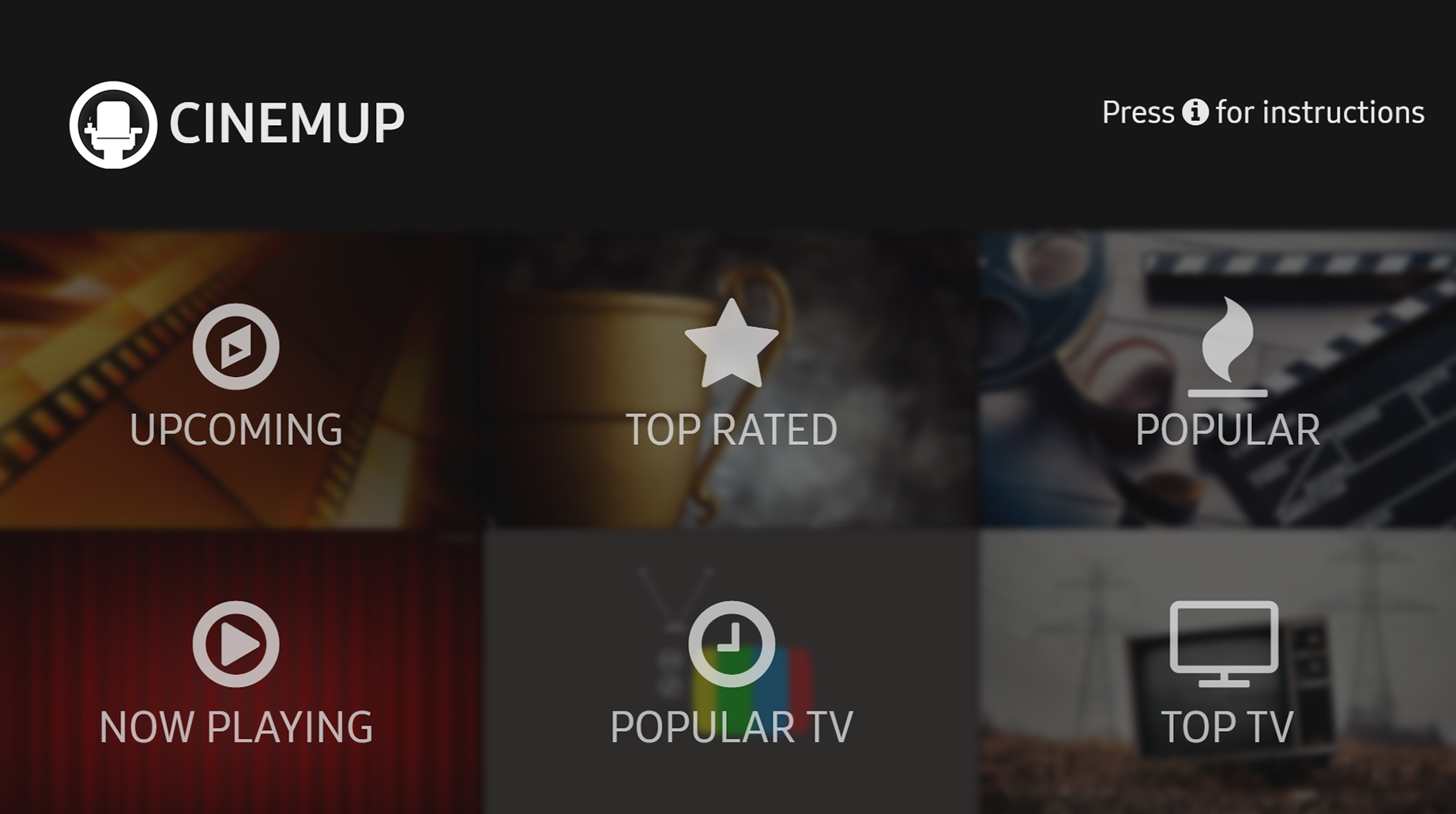}
                \vspace{-0.3cm}
		\caption{ \centering }
		\label{CineMup1}
    \end{subfigure}%
         \hfill
    \begin{subfigure}[b]{0.3\textwidth}
    	\includegraphics[width=\textwidth]					{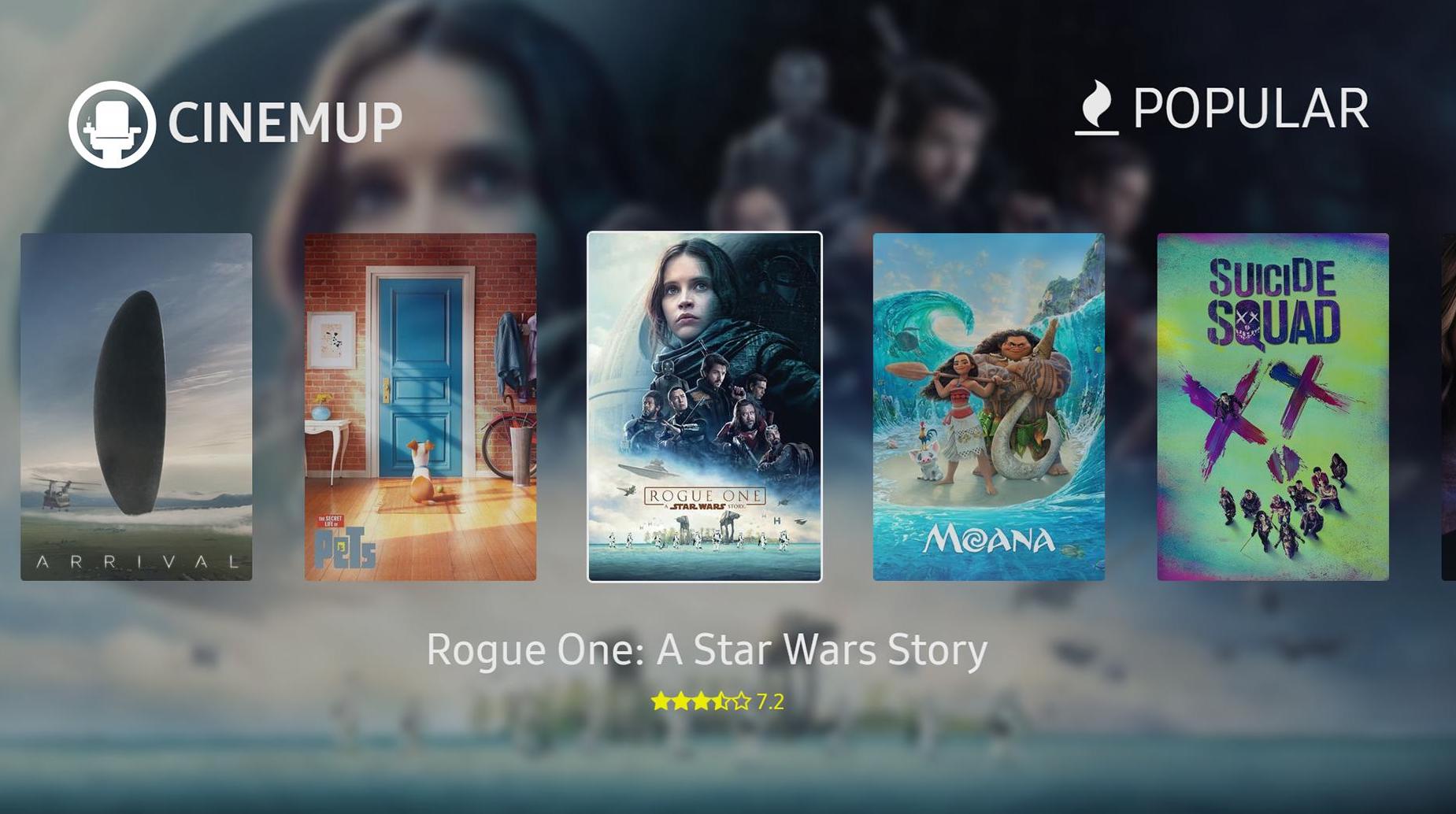}
                \vspace{-0.3cm}
        \caption{\centering }
        \label{CineMupEdgeNumber}
          \end{subfigure}
             \hfill
             \begin{subfigure}[b]{0.3\textwidth}
    	\includegraphics[width=\textwidth]					{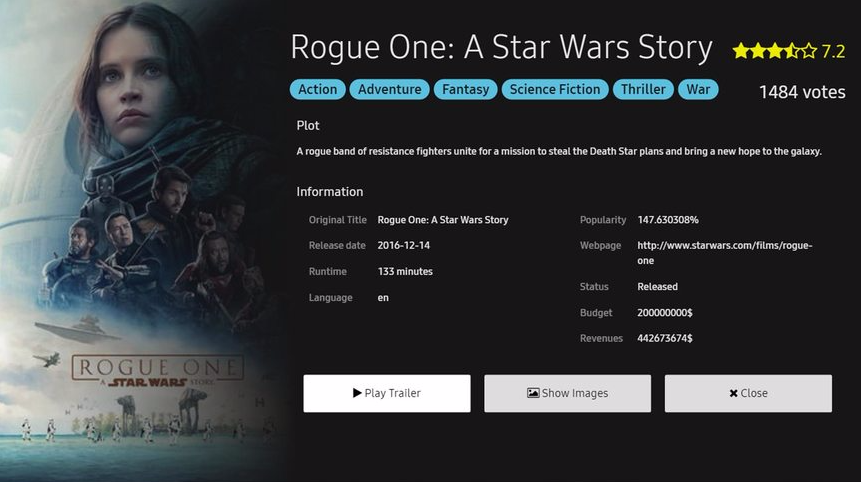}
                \vspace{-0.3cm}
        \caption{  \centering }
        \label{MemorygameEdgeNumber}
          \end{subfigure}
             \hfill
\caption{Three different screens for the CineMup smart TV app}
\label{Fig:CineMupLayout}
\end{figure*}

To generate a model for a smart TV app, it is necessary to know the different layouts that an app may follow. Although there is a freedom for the designers to design the layout per the user preferences, smart TV app has a limited number of layouts that have been constructed for the ease of use with the remote control device. For example, Fig. \ref{Fig:CineMupLayout} shows three different layouts presented in three different screens of the CineMup smart TV app used for case study later in this paper.

\begin{figure}
\centering
\includegraphics[width=2.5 in]				{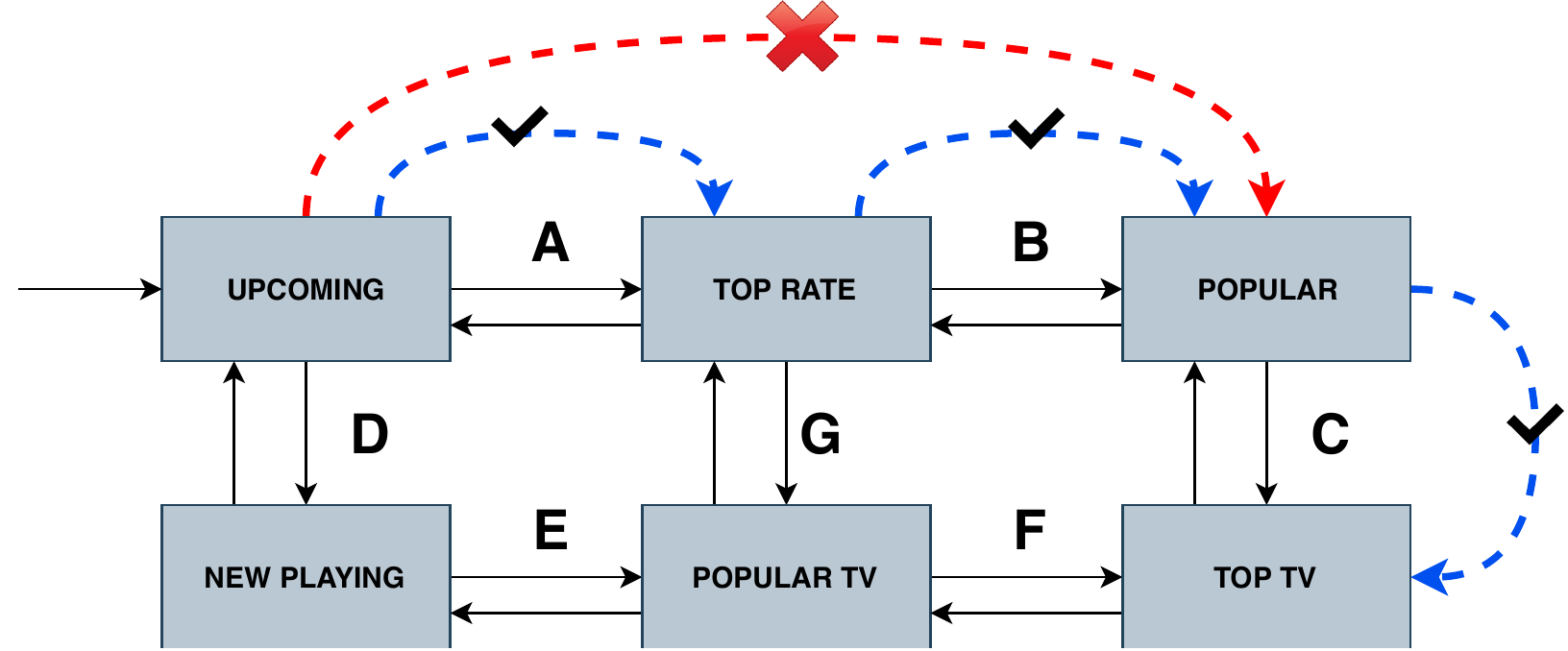}
\caption{Representation of the states in the Fig. \ref{CineMup1} with the right and wrong state transitions}
\label{Fig:rightandWrongNavigation}
\end{figure}

To navigate on the GUI of a smart TV app, the remote control device is the conventional way of interaction. The most critical issue here is the non-flexibility of the remote device to navigate from a widget to another due to its "one step" movement. For example, consider the different movement possibilities for the diagram in Fig. \ref{Fig:rightandWrongNavigation}. Here, this diagram represents the clickable widgets in Fig. \ref{CineMup1}. Take the "UPCOMING" button as a start widget for navigation. The navigation to the "POPULAR" button from the "UPCOMING" button in one step of the remote device is not possible because they are not adjacent. Hence, the user must pass through the "TOP RATE" button to reach the "UPCOMING" button.

Now, turning to the "TOP TV" button in Fig. \ref{Fig:rightandWrongNavigation}, there are three paths to reach this button from the "UPCOMING" button. Those path edges are [A:B:C], [A:G:F], and [D:E:F]. To reach "TOP RATE" from the "UPCOMING," for covering the [A] edge, the user must press the "Right" button on the remote device. In the same way, to cover the path [A:B:C], the user must press the remote device three times (i.e., Right, Right, then Down) to reach the "TOP TV" button. To construct a model that simulates this navigation process for the user, the model construction algorithms for mobile or desktop apps are useless here and there is a need for new algorithms to generate a new model.  

\section{The Model Generation Method}\label{ModelGenerationMethod}

In this section, we present our automated approach to generating a comprehensive model of the smart TV app. The model is generated with the help of our crawler that visits all the clickable states in the GUI of the smart TV app. Unlike the other crawling algorithms for desktop, web, and mobile apps, our crawling algorithm uses a full black-box crawling by simulating the actual interaction of the user with the app. Hence, there is no need to scan the app source code to generate the model. Also, following this approach, there is no need to repair the generated model for the redundant and obsolete states. By the end of the crawling, our strategy will generate the mega-model that forms a full representation of all widgets and transitions in states (nodes) and edges. We used the smart TV emulator in the Tizen SDK to run our crawler and mega-model generation algorithm. The following subsections give the detail of the crawling algorithm, the mega-model representation, and the notion of sub-model and model subtraction.

\subsection{The Smart TV App's Crawler and Its Black-box Mega-model Generation}

We have developed an algorithm to crawl the smart TV apps extensively using the remote control device emulator to discover the actually clickable widgets on the GUI. This crawling process aims to construct the mega-model that represents a comprehensive GUI model. We have adopted and developed the concepts used in the literature for desktop and mobile app crawling. In contrast to those approaches, we used the remote control device to explore the GUI directly without going to the details of the internal code structure of the app. Algorithm \ref{alg:Application-Creeper} shows the steps of the crawling method.

\begin{algorithm}
\scriptsize
\DontPrintSemicolon
\SetKwProg{Fn}{function}{:}{}
\SetKwFunction{FMain}{exporeTVApp}
\KwIn {$v_1$ is the starting or user selected state}
\KwIn {$It_{max}$ is the max number of crawling actions}
\KwOut {List of states to be modeled $L_{v}$ and the transitions among them}
Check the focus point \;
\If  {focus not set}{
        Display the message "Please Select the Focus Point"\;
}
Set $v_1$ as focus\;
Set $L_v$ as a new empty lists of states\;
\FMain{$v_1$,$It_{max}$,$L_v$}\;
\KwRet $L_v$\;
\;
\Fn{\FMain{$s$, $i$, $L_v$}}{
        \KwIn{$s$ is the starting state}
        \KwIn{$i$ is max number of actions still to perform}
        \KwIn{ $L_v$ is the list of visited states}
        \If(\tcc*[f]{max actions reached}){$i = 0$}{\KwRet\;}
        \ForEach {direction $d$ in $\{D_{Up}, D_{Down}, D_{Left}, D_{Right}, D_{OK}\}$}{
            apply direction $d$ to navigate to a new state $currentState$\;
            monitor emulator log for reaction\;
            \If {$currentState.active = true$}{
                Record the in/out transitions\;
                \If {$currentState$ $\not\in L_{v}$} {
                    add $currentState$ to $L_{v}$\;
                    \FMain{$currentState$, $i-1$, $L_v$}\;
                }}
        $Back$ to the parental state as $currentState$\;
        }
    }
\caption{\label{alg:Application-Creeper} The Steps of the Smart TV GUI Crawler Algorithm}
\end{algorithm}

As mentioned previously, the smart TV app technology depends mainly on the JavaScript programming language. The starting widget on a GUI of the app depends on the "focus" point specified by the developer in the JavaScript code. The focus point determines the location of the remote control cursor on the GUI. For the majority of the apps we explored, the focus point was not set during the development. Here, when the user opens the app, he must move the cursor using the remote control device to choose the focus point. To this end, our algorithm starts by checking the focus point. If the focus point was not set, then the algorithm will ask the tester to move the cursor and choose the focus point to start.

To explore the app comprehensively, from each newly discovered widget, the algorithm takes five directions $D_{Up} $, $D_{Down} $, $D_{Left} $, $D_{Right}$, and $D_{OK}$ to navigate on the GUI. When a new widget is discovered, the algorithm will add it to the final list of states that must be represented on the mega-model. The algorithm will continue this exploration until there are no new widgets to discover. The algorithm uses the $Back$ key for moving back to the parental widget and GUI window. Another stopping criterion for this crawling algorithm is the setting of the iteration number at the beginning. For some cloud-based smart TV apps (for example, YouTube), it is impossible to explore the app comprehensively. For example, in the case of YouTube, each video is a state on the model. Here, the list of videos is not finite and the algorithm will continuously discover new widgets. We set an iteration number in the algorithm for this situation. However, for precise and finite results in this study, we did not use such an app in the case studies.

\begin{figure}
\begin{centering}
\includegraphics[scale=0.24]{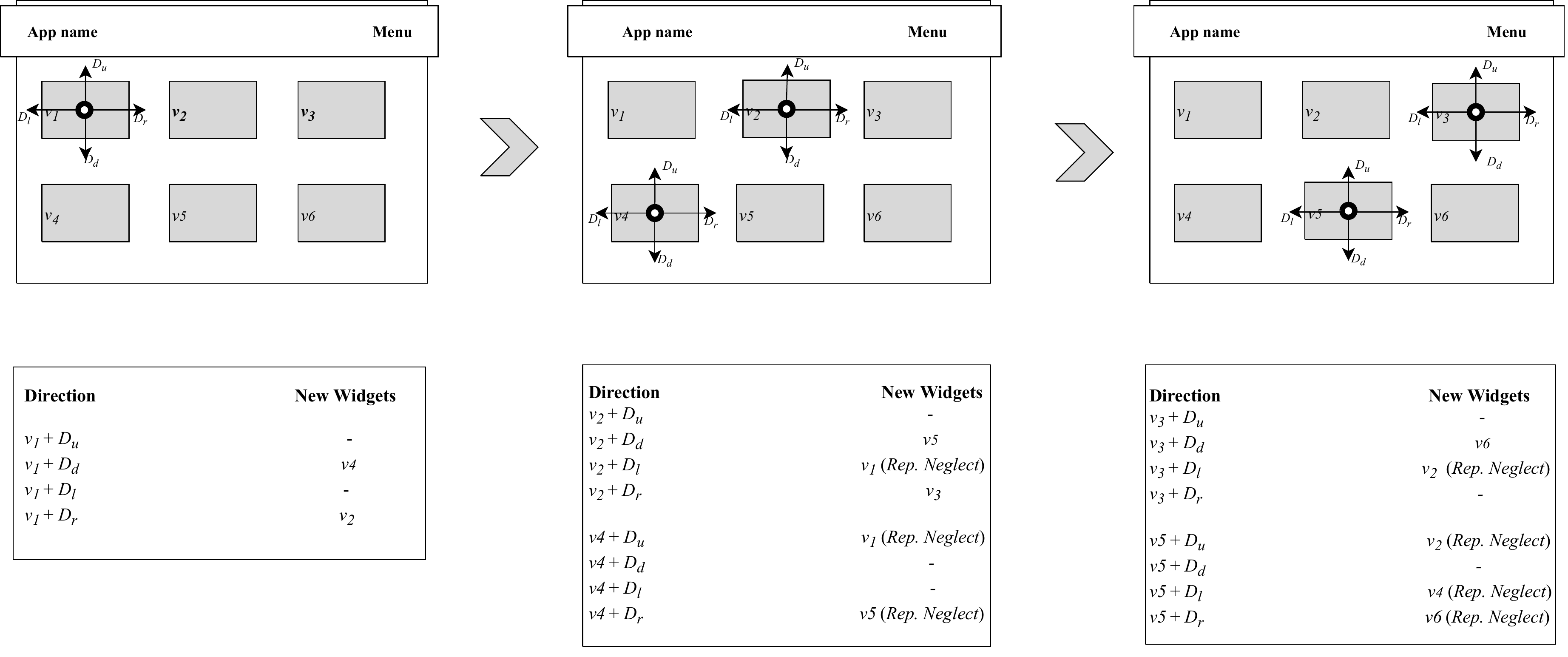}
\par\end{centering}
\caption{\label{fig:Proof-of-concepts} A Graphical Representation of the Smart TV GUI Crawling}
\end{figure}

To show the concepts of the crawling algorithm, we consider a running example for one window of the GUI with six widgets as shown in Fig. \ref{fig:Proof-of-concepts}. We consider $v_1$ as a starting widget. The algorithm starts from $v_1$ to crawl in four main directions $D_{Up}$ , $D_{Down}$ , $D_{Left}$ , and $D_{Right}$ for the full exploration of the GUI. For each direction, the algorithm checks for a new widget that can be discovered by listening to the log action of the smart TV emulator. Each new discovered clickable widget becomes a state on the mega-model. For example, from $v_1$, the up and left directions $D_{Up}$ , $D_{Left}$ will not lead to any widgets. The right direction $D_{Right}$ leads to $v_{2}$ and the down direction $D_{Down}$ leads to $v_{4}$. The algorithm then starts from the newly discovered widget, i.e., $v_{2}$. From $v_{2}$, the algorithm discovers the new states $v_{3}$ and $v_{5}$ in the $D_{Right}$ and $D_{d}$ directions. The algorithm also discovers $v_{1}$ in the $D_{Left}$ direction; however, it is neglected by the algorithm as it is already discovered. The algorithm will continue until it cannot find more widgets.

\begin{figure}
\begin{centering}
\includegraphics[scale=0.3]{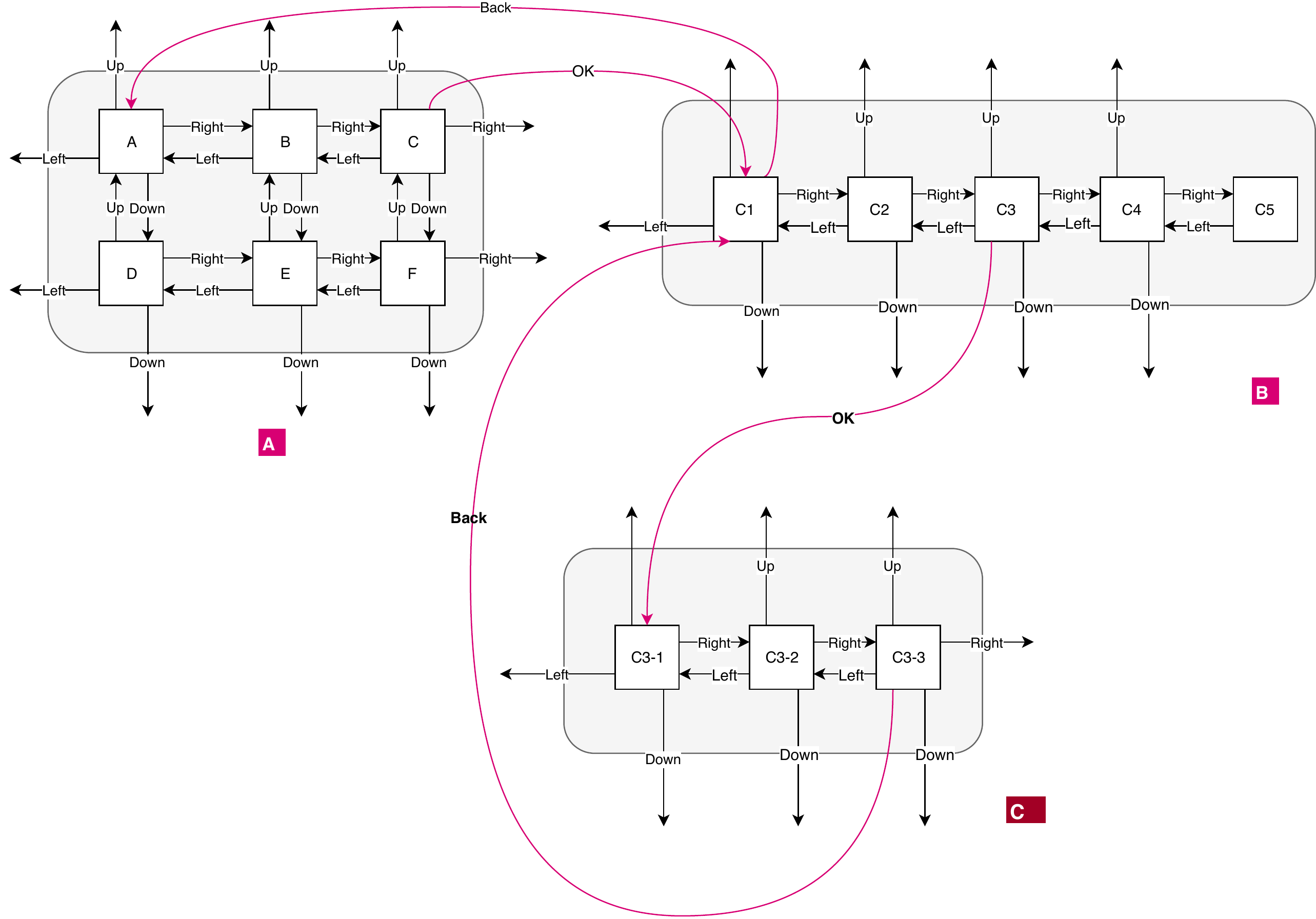}
\par\end{centering}
\caption{\label{fig:FullCrawling} Full crawling of the GUIs in Fig. \ref{Fig:CineMupLayout} }
\end{figure}

To consider all the directions by the algorithm to deeply discover an app with multiple windows, we also consider the $OK$ and $Back$ keys. Fig. \ref{fig:FullCrawling} shows a pilot deep crawling example for the smart TV app with three GUIs in Fig. \ref{Fig:CineMupLayout}.

We model the GUI of the smart TV app as a multigraph $\mathfrak{\mathcal{G}}=(N, E,s,t)$, where $N$ is a set of nodes, $N\neq\emptyset$, and $E$ is a set of edges. Here, $s:\,E\rightarrow N$ assigns each edge to its source node and $t:\,E\rightarrow N$ assigns each edge to its target node. One start node $n_{s}\in N$ is defined when the focus point set. The set $N_{e}\subseteq N$ contains the end nodes of $\mathfrak{\mathcal{G}}$, $N_{e}\neq\emptyset$. $N$ represents the widget elements of the GUI of the smart TV app. $E$ represent the possible transitions between the widget elements ($N$) using the remote control device. The widget here is the clickable widgets on the GUI. As previously mentioned, the crawler, for example, will not detect the label widgets on the GUI to be considered in the mega-model. By following this approach, we can avoid a great testing overhead time needed to repair the model and the obsolete generated test cases when all widget considered. Hence, the constructed mega-model represents a real user interaction with the app and all the test cases generated using this model can be executed directly.

\subsection{The Notion of Sub-model and Model Subtraction}

Due to the many paths and ways of reaching the widgets in the smart app and the transitions among them, most probably, the mega-model will contain many states and edges. The mega-model becomes a complicated model with many edges and nodes. To assure the successful execution of a generated test case, it must start from the starting state of the graph and goes to a specific end state. Here, a test case is a sequence of steps to pass several nodes and paths to reach a final destination that usually leads to an action on the app. Hence, a step is a node or a state on the mega-model graph and the edge between two nodes is the move action by the user with the remote control device.

As the end nodes are representing the final destination for some action on the app, a natural way to generate a set of test cases is to consider all the end destination nodes on the mega-model and generate paths from the start to the end nodes. Although feasible, this test generation approach is exhaustive and computationally expensive. Heuristic search-based algorithms is an approach to generate and optimize test cases. However, it is more suitable to test the whole app not a specific function on the app. Besides, generating all the possible paths for the mega-model will lead to run many impractical test cases which may not be needed for testing a specific function on the app. Also, the search-based algorithm becomes complicated and computationally expensive for large mega-models and there is a need for specially designed data structures.

\begin{figure}
\centering
\includegraphics[width=3.3 in]				{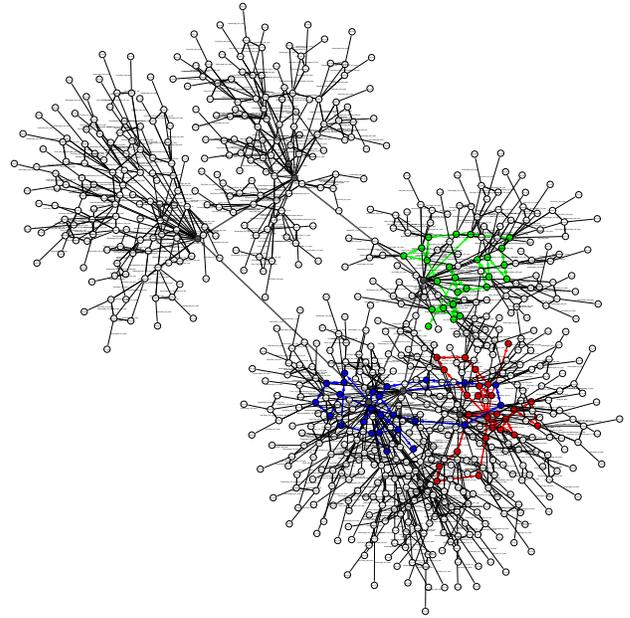}
\caption{The Mega-model with Sub-models Highlighted for the CineMup Smart TV app}
\label{Fig:MegaModelWithSubGraphs}
\end{figure}

In this paper, we introduce the notion of sub-model to generate the test cases. Here, we generate the test cases based on the selected functions that we want to test. A function is an expected output after running a sequence of steps on the app. For example, the "Play Trailer" in Fig. \ref{MemorygameEdgeNumber} is a destination node in the app that is used for "play" function of a specific movie trailer track. Following this approach, we can test more specific functions and generate more effective test cases. Also, as the models are sub-models and they are relatively simpler than the mega-model, we can generate the test cases with less computation time and resources. As an example, We identified thee sub-models on the mega-model in Fig. \ref{Fig:MegaModelWithSubGraphs}.

\begin{figure*}
	\centering
	\begin{subfigure}[b]{0.32\textwidth}
		\includegraphics[width=\textwidth]					{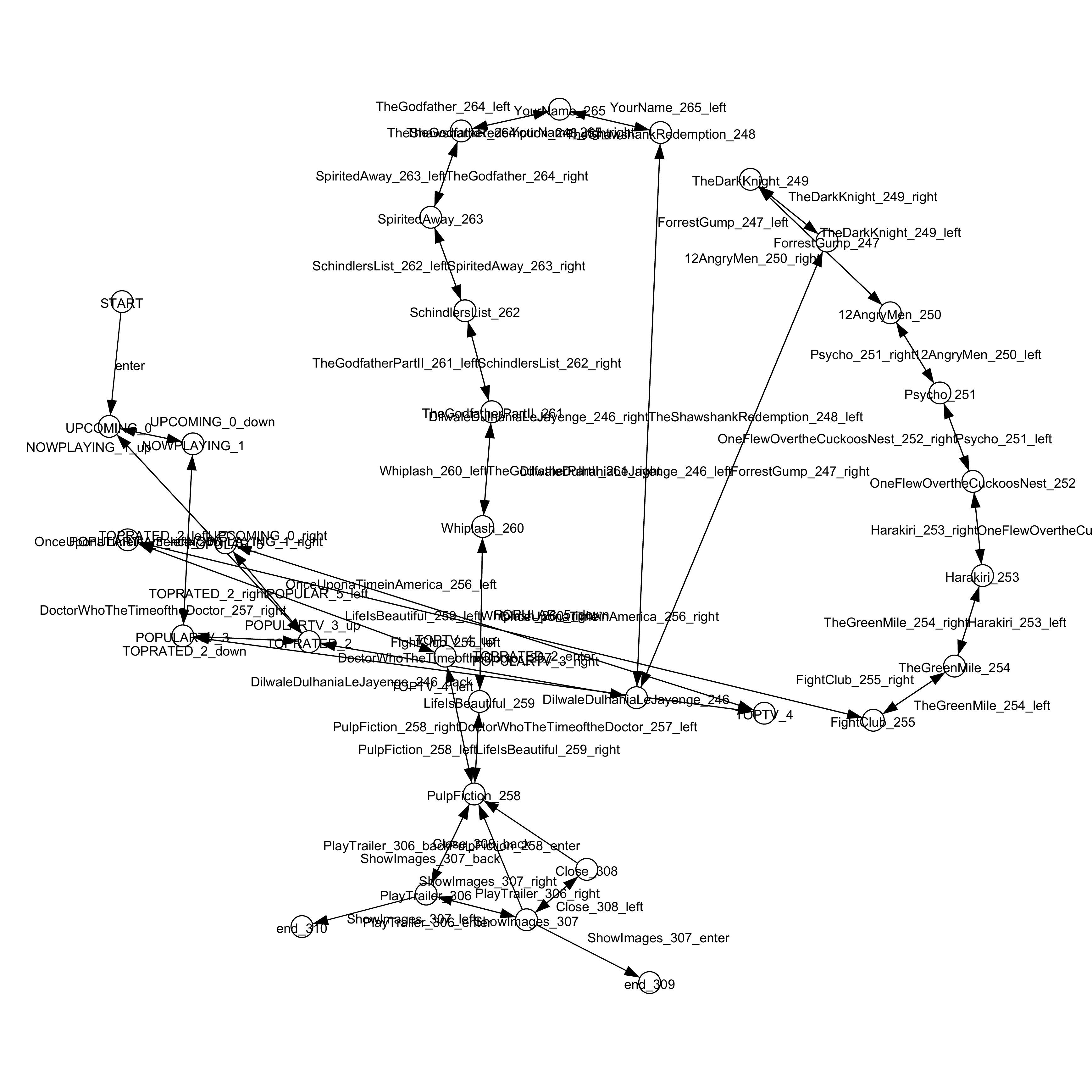}
                \vspace{-0.7cm}
		\caption{Sub-model (Green)}
		\label{GreenGraph}
    \end{subfigure}%
         \hfill
    \begin{subfigure}[b]{0.32\textwidth}
    	\includegraphics[width=\textwidth]					{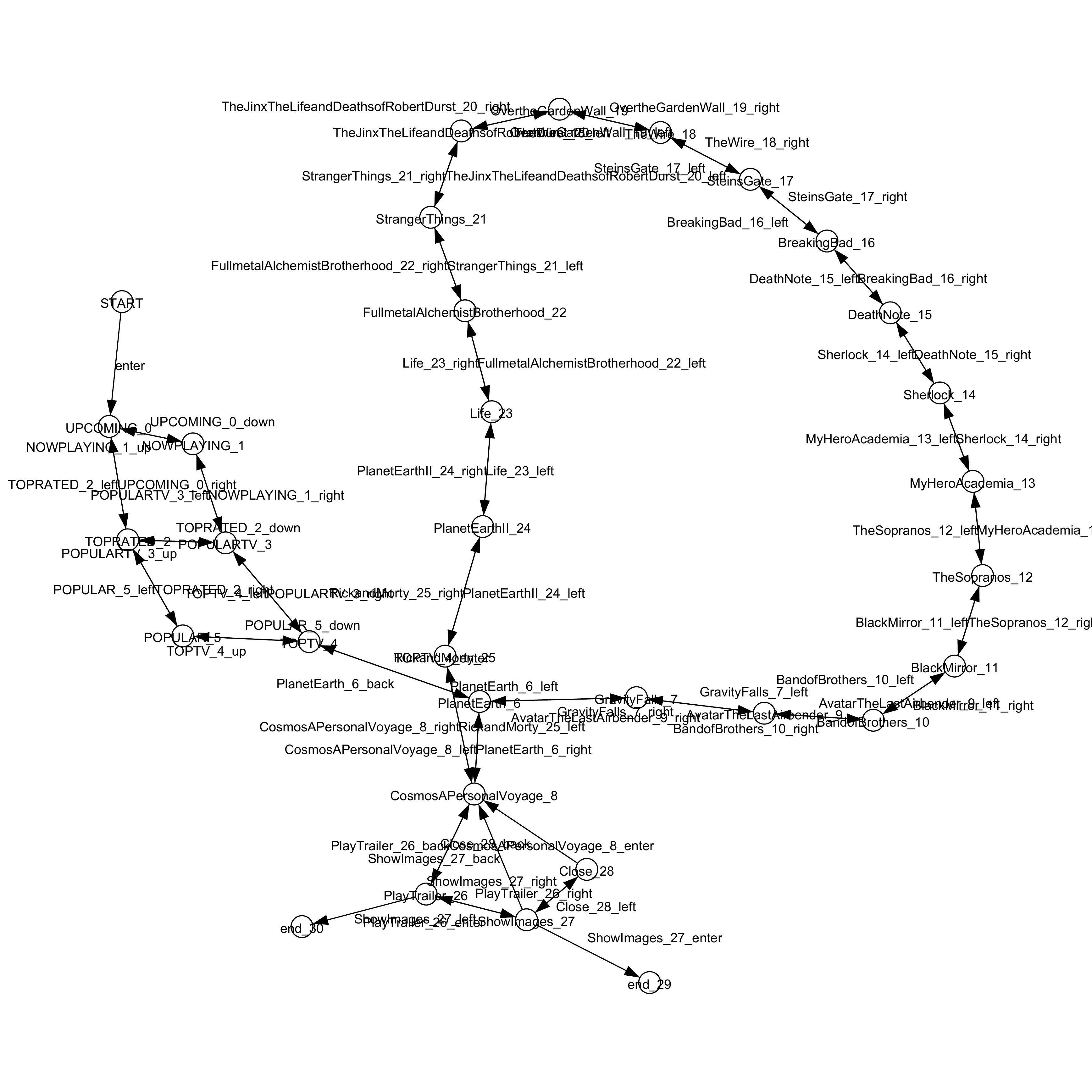}
                \vspace{-0.7cm}
        \caption{Sub-model (Red)}
        \label{fig:decisioncoverageRand1}
          \end{subfigure}
     \begin{subfigure}[b]{0.32\textwidth}
    	\includegraphics[width=\textwidth]					{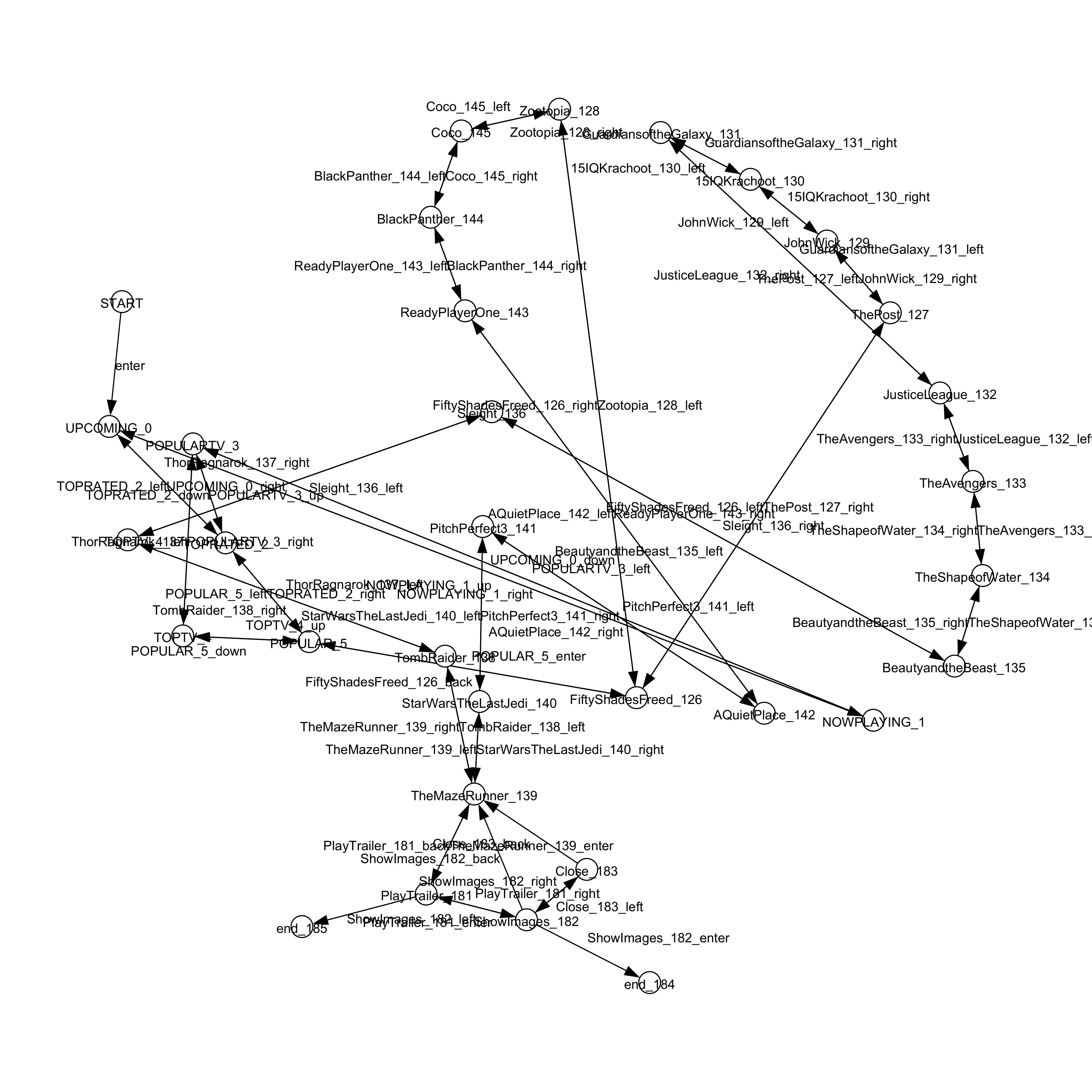}
                \vspace{-0.7cm}
        \caption{Sub-model (Blue)}
        \label{fig:decisioncoverageRand2}
          \end{subfigure}     
\caption{The extracted sub-models represting the colored graphs illustrated in Fig. \ref{Fig:MegaModelWithSubGraphs} for the CineMup Smart TV app }\label{fig:SubModelscolored}
\end{figure*}

As shown in Fig. \ref{Fig:MegaModelWithSubGraphs}, we highlighted the sub-models on the mega-model in different colors (Red, Blue, and Green). Each color is a directed graph with start and end nodes that represent a complete model with all related nodes and edges to fulfill a specific action. Our testing strategy separates these sub-models based on the user selection of the end nodes. For example, Fig. \ref{fig:SubModelscolored} shows the graphs for the (Red, Blue, and Green) sub-models in Fig. \ref{Fig:MegaModelWithSubGraphs}. The following section shows the test generation and execution algorithm for the sub-models.

\section{The Test Generation Strategy}\label{TestGenerationMethod}

To generate the test cases for the smart TV app from the sub-model graph, we introduce  a testing criteria, \textit{All Edge Coverage}, that requires to cover all the edges between the nodes to execute all transitions from a widget to another. A test case $t$ is a sequence of nodes $n_{1},n{}_{2},..,n_{n}$, with a sequence of edges $e_{1},e{}_{2},..,e_{n-1}$, where $e_{i}=(n_{i,}n_{i+1})$, $e_{i}\in E$, $n_{i}\in N$. Further on, $n_{1}=n_{s}$, and $n_{n}\in N{}_{e}$. The Test Set $T$ is a set of test cases. $T$ achieves the \textit{All Edge Coverage} criterion if each edge $e\in E$ is traversed at least one by a test case in $T$. The Algorithm \ref{alg:GenerateTestCases} shows the pseudocode of the used test generation algorithm.

\begin{algorithm}
\scriptsize
\KwIn{A sub-model Graph ($\mathcal{G}$)}

\KwOut {The test set $T$}

$T$ $\leftarrow$ $\emptyset$, $COVER$ $\leftarrow$ $\emptyset$, $ALL \leftarrow \emptyset$

\For(\tcp*[h]{build the $COVER$ set with all the edges to be covered}){$e \in E$}{
    let  $n_{1}$ and $n_{2}$ be the head and tail of $e$, where \{$n_{1}, n_{2}\} \in N$, $e\in E$\;
    $COVER$ $\leftarrow$ $COVER$ $\cup$ \{($n_{1}$, $e$, $n_{2}$)\} 
}
\For(\tcp*[h]{build all the useful paths that can be traversed}){$c \in COVER$ }{
        Let $z$ be a path in $\mathfrak{\mathcal{G}}$ starting with $n_{s}\in N$ and ending with any $n_{e}\in N_{e}$ that covers $c$\;
        $ALL$ $\leftarrow$ $ALL$ $\cup$ \{$z$\}\;
}
\While(\tcp*[h]{select the best paths as tests}){$COVER\neq\emptyset$} {
    $best$ $\leftarrow$ a path from $ALL$, which contains maximal number of elements from $COVER$ as its sub-paths \;
    $T$ $\leftarrow$ $T\cup best$\;
    Let $x$ be the set of edges of $best$\;
    $COVER$ $\leftarrow$$COVER$ /$x$
}

\caption{\label{alg:GenerateTestCases}The test generation algorithm from the sub-model}
\end{algorithm}

As an input, the test generation algorithm takes the subtracted sub-model $\mathcal{G}$ of the smart TV app. We have adopted and deployed our previously developed Prioritized Process Test (PPT) algorithm for the test generation from the model in Algorithm \ref{alg:GenerateTestCases}. We have extensively evaluated and assessed the PPT algorithm in other published studies, e.g., \cite{bures2017prioritized}. In contrast to the PPT algorithm, the Algorithm \ref{alg:GenerateTestCases} reflects no edge priorities and is simplified to generate $T$ with \textit{All Edge Coverage} level only. At the start of the algorithm, edges which shall be present in the test cases of $T$ are expanded by the head and the tail nodes and added to the set $COVER$. The set $COVER$ works as a stack containing the edges not yet covered by the generated test cases. The $COVER$ size reduced during the subsequent iterations of the algorithm by deleting the already covered test cases. As the next step during the initialization, all the possible paths in $\mathcal{G}$, starting from the $n_{s}\in N$ and ending in any node $n_{e}\in N_{e}$, which contain a path $p\in P$ are collected in $ALL$. These paths in $ALL$ represent the candidates for the test cases in $T$. Similarly, the set $ALL$ is reduced during the subsequent iterations of the algorithm. During the whole cycle, best test paths from $ALL$ are incrementally added to $T$. Here, from $ALL$, the algorithm selects the paths, which contain the maximum number of elements from $COVER$ and put them as the final test suite in $T$. In each iteration, the algorithm removes the covered elements from $COVER$. 

As mentioned earlier, the output of the test generation algorithm is a test suite that contains a set of test cases. Each test case is a sequence of steps towards a destination on the smart TV app to fulfill a function. Typically, a test case here is a long sequence of steps to be executed on the app. The length of the test case depends on the depth of the destination function on the graph model and how far the destination node is from the starting point.

\section{A Comprehensive Testing Strategy}\label{TheTestingStrategy}

\begin{figure}
\centering
\includegraphics[width=3.3 in]				{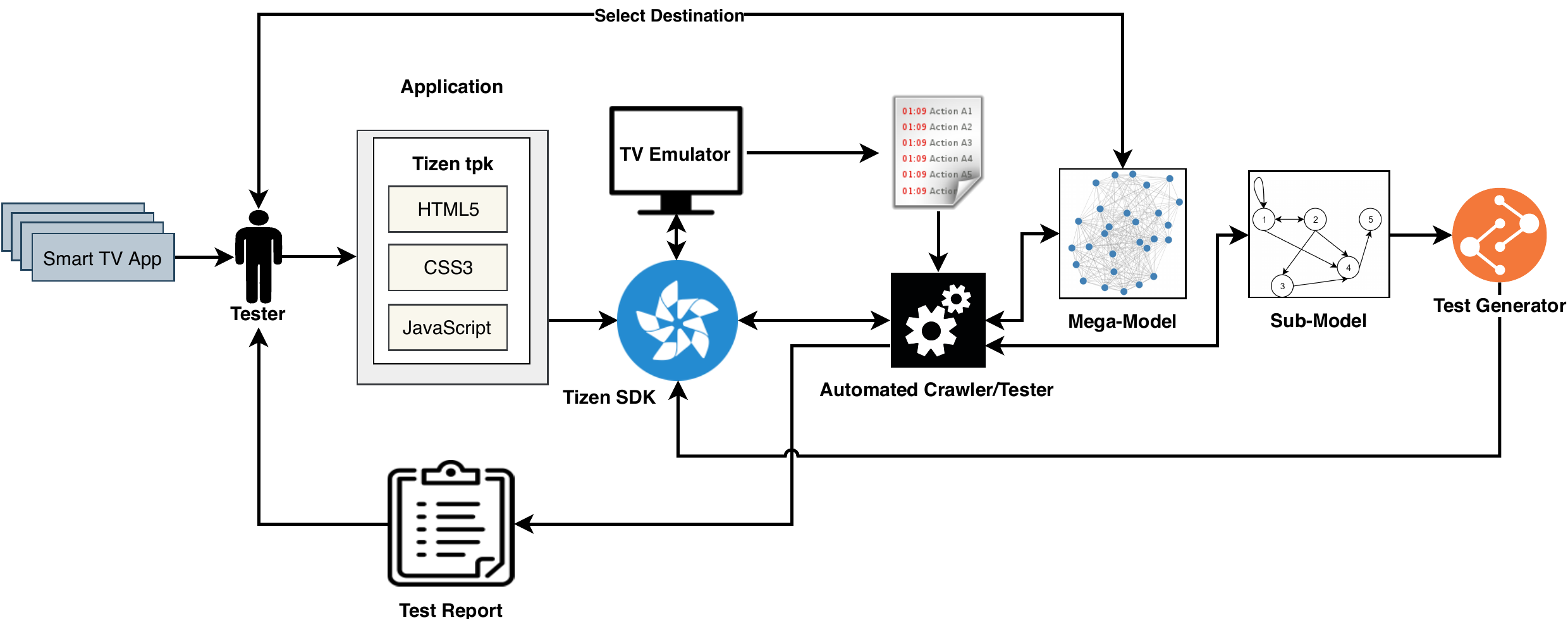}
\caption{Testing Framework for Smart TV apps}
\label{Fig:TestingFramework}
\end{figure}

We have implemented our testing strategy in an automated testing framework called EvoCreeper within the Tizen SDK. Fig. \ref{Fig:TestingFramework} shows a complete diagram of the framework, including the essential testing components that we have illustrated so far. The detail of the crawler implementation and the full source is available for download\footnote{Smart TV crawler download page https://github.com/bestoun/EvoCreeper}.

The tester installs the Tizen app (with tpk extension) on the smart TV emulator which is included in the SDK. If the source code of the app is available, it can also be installed through the SDK. The interaction with the app can be monitored by a real-time log file that is accompanied by the emulator. Our crawler algorithm examines each widget on the app using the emulator and the remote device. The automated tester then starts to create the mega-model based on the interaction output in the log file. When the crawler finishes its crawling on the app, the mega-model is presented to the tester. The tester then chooses the destinations that he wants to test on the mega-model. Based on the selected destination, the automated tester then will create the sub-model. The sub-model is then used to generate the test cases using Algorithm \ref{alg:GenerateTestCases}. We have implemented the test generator algorithm from the sub-model in the Oxygen environment\footnote{The web page of the oxygen project http://still.felk.cvut.cz/oxygen/}. 

Once the test cases are generated by our test generation algorithm, the framework will execute each test case on the emulator and monitor the output of each one of them. Here, we consider a test case to pass when it can be run without any interruption from a start to the end on the emulator. We also considered a test case as "fail" when it would not provide the expected output. To assess the effectiveness of the test cases, there is a need to inject faults into the app and evaluate how the test cases can detect them, which lead to a classical mutation testing process. The following Section gives the detail of this process.


\section{Smart TV app's Mutation Operators}\label{MutationOperators}

An essential contribution of this paper is the proposed mutation operators that we designed for smart TV apps. Mutation analysis is a well-known method to assess the effectiveness of the generated test cases by seeding faults into a program and then try to detect those seeded faults \cite{PAPADAKIS2018}. Mutation analysis uses mutation operators to seed the faults into the program. Designing mutation operators for different kind of apps and programming languages is an active research area. For example, mujava \cite{Ma2005} for java language mutation or the mutation of Android apps \cite{Deng2017}. A well-designed mutation operator leads to effective test cases. In the literature, there are two approaches to design mutation operators. The first approach is to rely on some well-defined faults models to design the operators that in turn, leads to define each mutant. The classical mujava tool is a well-known example of this approach that used for class level mutation of java programs. The second approach is to analyze the language that the app is written to identify the syntactic elements that can be modified to simulate a mistake that a typical program may face.

Although many mutations operators and tools can be used for different kind of programs and apps, none of them can be used for smart TV app testing. The whole smart TV app testing has not been addressed in the literature. To this end, we have designed a set of mutants for each case study we used during the testing. The operators are based on the common faults that the programmer may make during the programming. Table \ref{MutantTable} shows the list of mutation operators we designed for the smart TV testing. The Table also shows the acronym of the operator and an example of the code being mutated.

\begin{table*}
\caption{Smart TV mutation operators with example}\label{MutantTable}
\scriptsize
\begin{centering}
\begin{tabular}{|p{32mm}|c|l|}
\hline 
Mutation Operator & Acronym & Example\tabularnewline
\hline 
\hline 
\multirow{2}{*}{Wrong Address Request } & \multirow{2}{*}{RAR} & \textbf{Original:} getHttpRequestImages(\textquotedbl{}https://api.themoviedb.org/\textquotedbl{});\tabularnewline
\cline{3-3} 
 &  & \textbf{Mutant:} getHttpRequestImages(\textquotedbl{}https://api.themoviedb.org/3/\textquotedbl{}); \tabularnewline
\hline 
\multirow{2}{*}{Non-existing Element} & \multirow{2}{*}{NEE} & \textbf{Original:} document.getElementById('trailer').play();\tabularnewline
\cline{3-3} 
 &  & \textbf{Mutant:} document.getElementById('trai').play();\tabularnewline
\hline 
\multirow{2}{*}{Non-existing Attribute} & \multirow{2}{*}{NEA} & \textbf{Original: }card2.lastChild.setAttribute(\textquotedbl{}class\textquotedbl{},
\textquotedbl{}card-front\textquotedbl{});\tabularnewline
\cline{3-3} 
 &  & \textbf{Mutant: }card2.lastChild.setAttribute(\textquotedbl{}clas\textquotedbl{},
\textquotedbl{}card-front\textquotedbl{});\tabularnewline
\hline 
\multirow{2}{*}{Non-existing Feature} & \multirow{2}{*}{NEF} & \textbf{Original:} \$(\textquotedbl{}.film-list\textquotedbl{}).slick('slickNext');\tabularnewline
\cline{3-3} 
 &  & \textbf{Mutant:} \$(\textquotedbl{}.film-list\textquotedbl{}).slick('slickNe');\tabularnewline
\hline 
\multirow{2}{*}{Null Variable Replacement} & \multirow{2}{*}{NVR} & \textbf{Original:} starsEmpty = 5 - starsFilled - starsSemiFilled; \tabularnewline
\cline{3-3} 
 &  & \textbf{Mutant:} starsEmpty = null; \tabularnewline
\hline 
\multirow{2}{*}{Wrong Calculation} & \multirow{2}{*}{WRC} & \textbf{Original:} index = index -MAX;\tabularnewline
\cline{3-3} 
 &  & \textbf{Mutant:} index = index -MAX{*}6;\tabularnewline
\hline 
\multirow{2}{*}{Badly Assigned Variable} & \multirow{2}{*}{BAV} & \textbf{Original:} index-{}-;\tabularnewline
\cline{3-3} 
 &  & \textbf{Mutant:} MAX-{}-;\tabularnewline
\hline 
\multirow{2}{*}{Incorrect Function Call} & \multirow{2}{*}{IFC} & \textbf{Original:} changePage(index);\tabularnewline
\cline{3-3} 
 &  & \textbf{Mutant:} changePag(index);\tabularnewline
\hline 
\multirow{2}{*}{Non-existent Event} & \multirow{2}{*}{NXE} & \textbf{Original:} document.body.removeEventListener(\textquotedbl{}keydown\textquotedbl{},handelPageOne,false); \tabularnewline
\cline{3-3} 
 &  & \textbf{Mutant:} document.body.removeEventListener(\textquotedbl{}\textquotedbl{},handelPageOne,false); \tabularnewline
\hline 
\end{tabular}
\par\end{centering}
\end{table*}

As it is clear from Table \ref{MutantTable}, the mutants are simulating artificial faults that can be injected into the smart TV apps. The mutants are throwing excepts in the program when they are detected. The mutants are simulating different faulty scenarios, such as incorrectly specified HTML element mutants, wrong address when querying the server, and variable names that are discovered when the app runs. We have injected the case study apps used in this paper with these faults manually and then we tried to detect them automatically. The following section shows the results of this evaluation process. 

It should be mentioned here that developing an automated mutation testing tool for smart TV app is out of the scope of this paper. However, it could form an active research direction in the future. To best of our knowledge, this study is the first study towards this direction.

\section{Empirical Evaluation}\label{Evaluation}

To evaluate our newly developed approach, we used an automated framework to deploy three case studies of smart TV apps for testing. We first used Our automated framework to reverse engineer the three smart TV apps to generate the mega-model for each one of them. From this mega-model, we select several sub-models that represent possible user preferred or targeted testing paths based on the destination node on the mega-model. Then we used the sub-models to generate the test cases. Based on the mutation generation approach that we presented earlier, we have generated several mutants for the subjected smart TV apps in this empirical evaluation. To implement and evaluate these steps, our empirical evaluation comprises essential phases: model generation efficiency, and test case effectiveness and mutation testing. All of these phases were done on Tizen Studio and its TV emulator that was installed on MacBook Pro PC with macOS High Sierra, 2.9 GHz Intel Core i5 processor and 8GB of memory.  

\subsection{Empirical Subjects }

The research in the smart TV app testing is in its early stage as it is relatively a new technology. Due to its new software development environment, the number of developed smart TV apps is much less than mobile apps for example. As mentioned, our goal is to create the mega-model and extract the submodel for test generation. For the small and straightforward apps, it may be feasible to test them manually. Even if we use our framework to test those apps, there will be a small model graph with a few nodes and edges that may not reflect the capabilities of our developed framework. However, the framework should be effective for small, medium, and large size apps. To this end, we choose three apps with small, medium, and large sizes (in term of number of widgets) for our empirical evaluation. Table \ref{AppObjectsForExperiments} shows the name and source of each app.

\begin{table}
\centering
\caption{Apps used in the case study}\label{AppObjectsForExperiments}
\scriptsize
\begin{tabular}{|c|l|>{\raggedright}p{5cm}|}
\hline 
ID & App & Source\tabularnewline
\hline 
\hline 
1 & CineMup & https://github.com/daliife/Cinemup\tabularnewline
\hline 
2 & ChessLab TV & http://tiny.cc/qd5igz\tabularnewline
\hline 
3 & Memory game & http://tiny.cc/gi5igz\tabularnewline
\hline 
\end{tabular}

\end{table}

We choose three well-developed smart TV app with small, medium, and large sizes. Those apps are CineMup, ChessLab TV, and Memory game. The CineMup is a large size smart TV app used to search for different types of movies and shows. These shows and movies are categorized in the app with the data and a short description of the movies along with the trial of it. ChessLab is a medium sized smart TV app designed to improve the players' skills of the chess game by learning patterns, solutions, and tactics used before. Whereas, the Memory, is a small size smart TV app used as a puzzle game for memorization.

\subsection{Model Generation Efficiency}

\begin{figure*}
	\centering
	\begin{subfigure}[b]{0.3\textwidth}
		\includegraphics[width=\textwidth]					{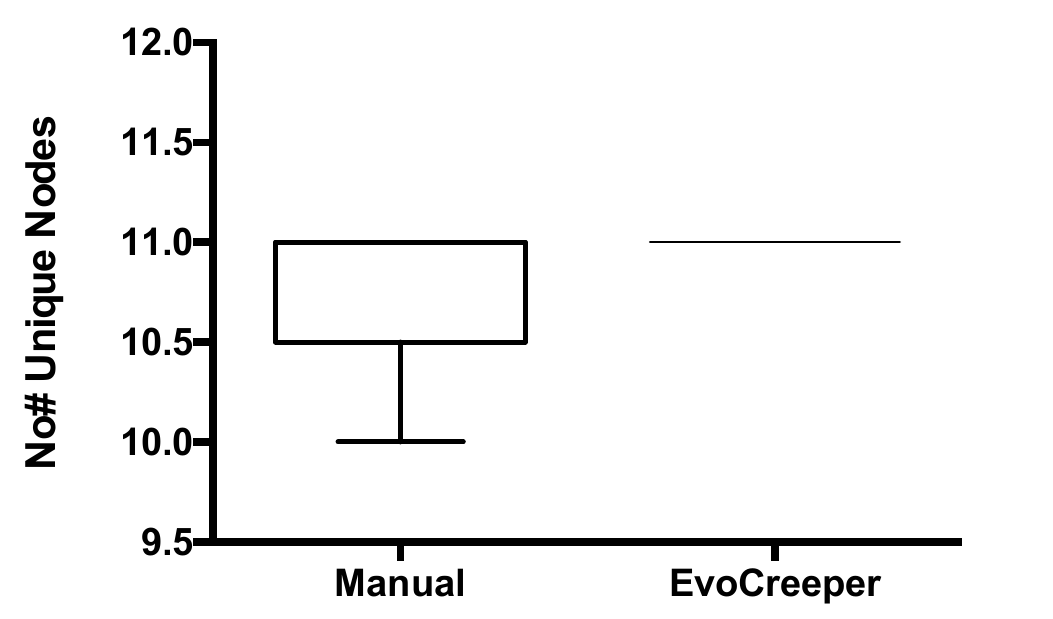}
                \vspace{-0.3cm}
		\caption{ChessLab \centering }
		\label{ChessLabNodeNumber}
    \end{subfigure}%
         \hfill
    \begin{subfigure}[b]{0.3\textwidth}
    	\includegraphics[width=\textwidth]					{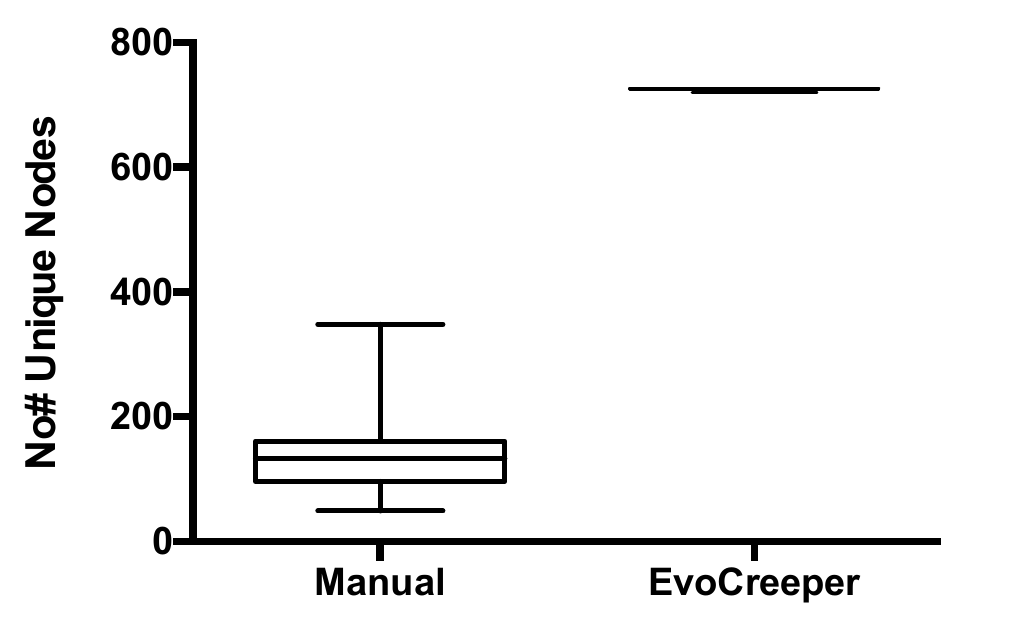}
                \vspace{-0.3cm}
        \caption{CineMup \centering }
        \label{CineMupNodeNumber}
          \end{subfigure}
             \hfill
             \begin{subfigure}[b]{0.3\textwidth}
    	\includegraphics[width=\textwidth]					{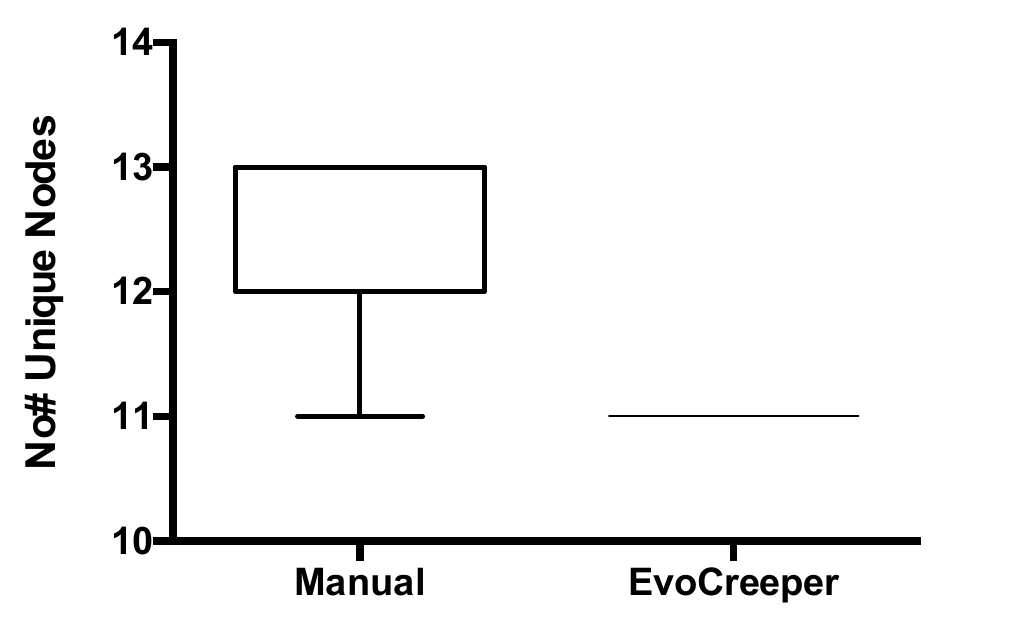}
                \vspace{-0.3cm}
        \caption{Memory game  \centering }
        \label{MemorygameNodeNumber}
          \end{subfigure}
             \hfill
\caption{Comparing the number of unique nodes detected by the automated crawler with the manual exploration}
\label{NumberOfUniqueNodeBoxPlot}
\end{figure*}

\begin{figure*}
	\centering
	\begin{subfigure}[b]{0.3\textwidth}
		\includegraphics[width=\textwidth]					{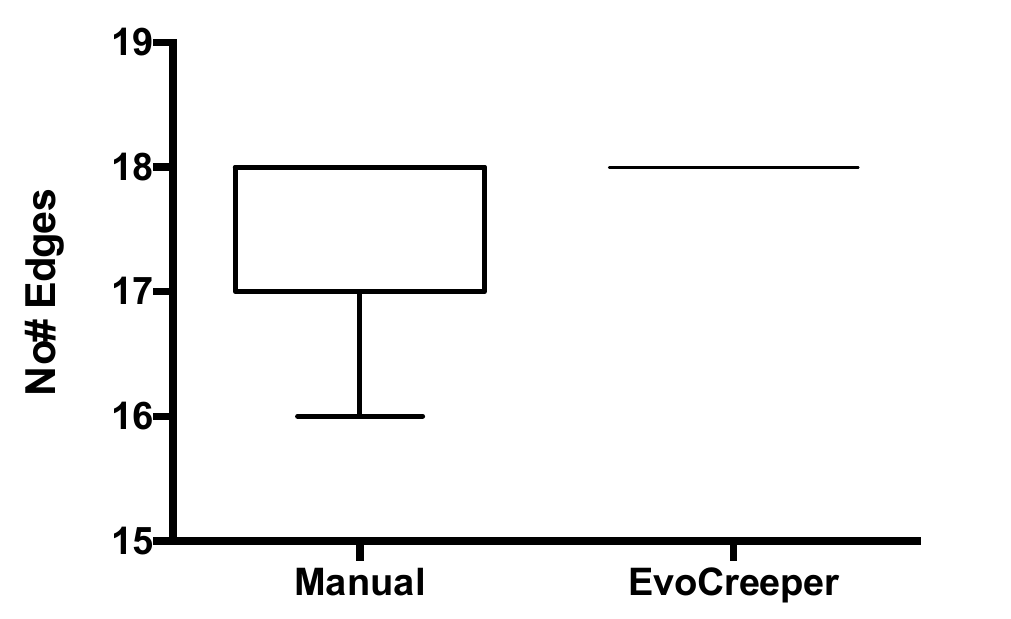}
                \vspace{-0.3cm}
		\caption{ChessLab \centering }
		\label{ChessLabEdgeNumber}
    \end{subfigure}%
         \hfill
    \begin{subfigure}[b]{0.3\textwidth}
    	\includegraphics[width=\textwidth]					{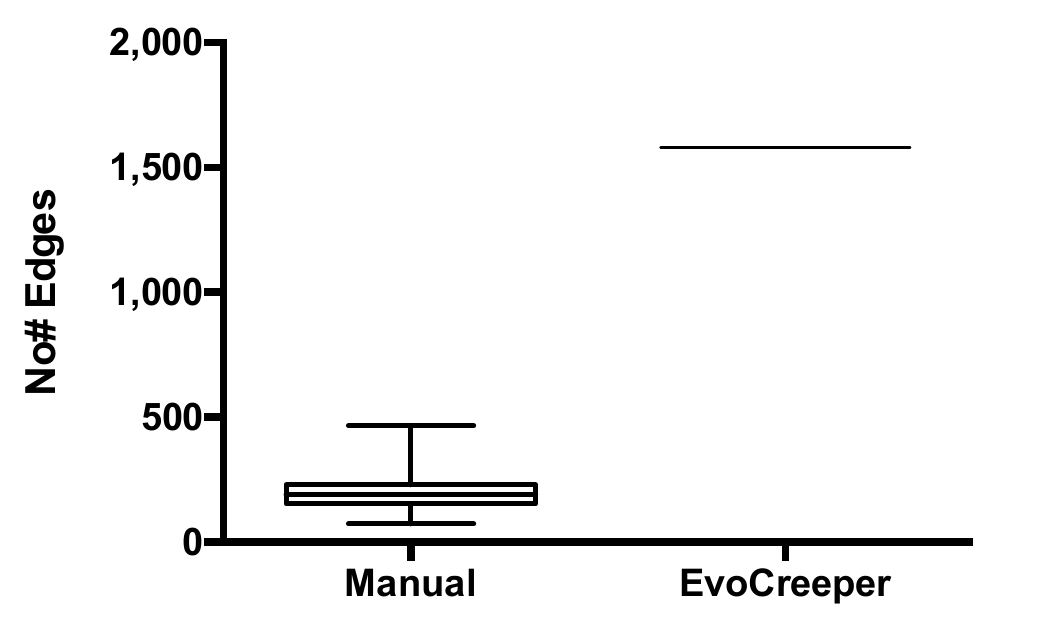}
                \vspace{-0.3cm}
        \caption{CineMup \centering }
        \label{CineMupEdgeNumber}
          \end{subfigure}
             \hfill
             \begin{subfigure}[b]{0.3\textwidth}
    	\includegraphics[width=\textwidth]					{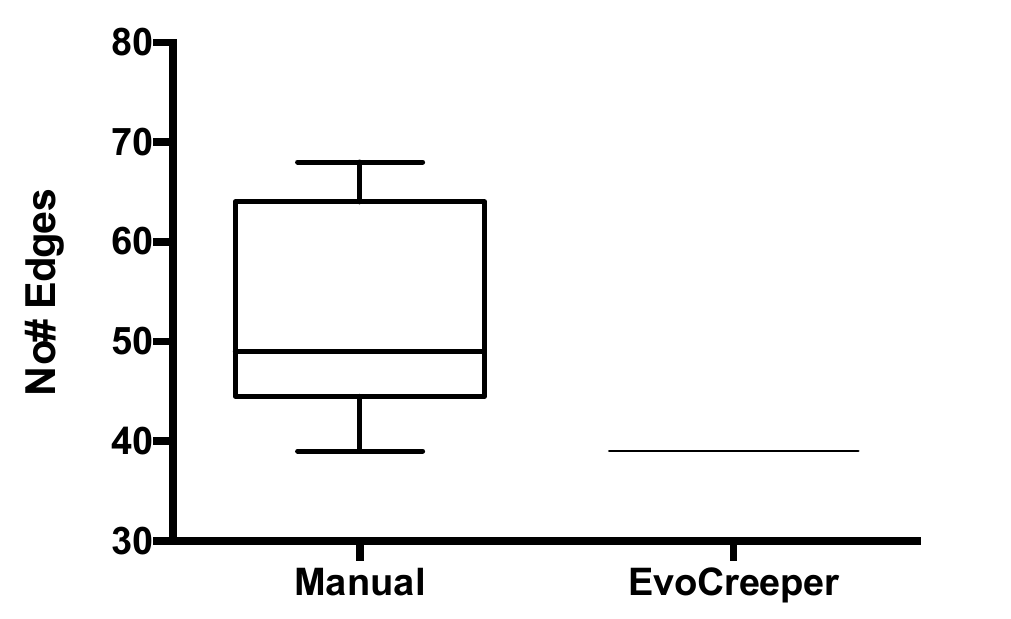}
                \vspace{-0.3cm}
        \caption{Memory game  \centering }
        \label{MemorygameEdgeNumber}
          \end{subfigure}
             \hfill
\caption{Comparing the number of Edges detected by EvoCreeper with the manual exploration}
\label{NumberOfUniqueEdgesBoxPlot}
\end{figure*}

In this experiment, we aimed to assess the efficiency of our model generation strategy as compared to manual exploration. We also aim to manually check the generated models. To achieve this aim, we compared the number of unique widgets (states on the model) explored and detected by our strategy as compared to the manual exploration of the app. To conduct the manual exploratory testing, we formed a group of 60 students from the software quality class into four groups. Each group provided and trained with a software package with Tizen studio and emulator that can be run on notebooks or desktop computers. We modified the remote control module of the emulator to record the states exploration history of the students for each smart TV app. The recorded information consisted of keys pressed, widgets press, and the transitions performed by the students. Besides, the time of the exploration is recorded for each student. To avoid mistakes and overhead time, we instructed each group of students for how to explore and use our experimental packages. We also instructed the students to stop the exploration recorder when he thinks that he has examined the smart TV app completely. We compared the number of the unique states and edges discovered by our automated crawler with manual exploration. Fig. \ref{NumberOfUniqueNodeBoxPlot} and \ref{NumberOfUniqueEdgesBoxPlot} show the comparison results. To show the statistical values and fair comparison in the different records by the students and the non-deterministic results obtained by them, we used box plots and whisker to show the comparison.

\begin{figure*}
	\centering
	\begin{subfigure}[b]{0.3\textwidth}
		\includegraphics[width=\textwidth]					{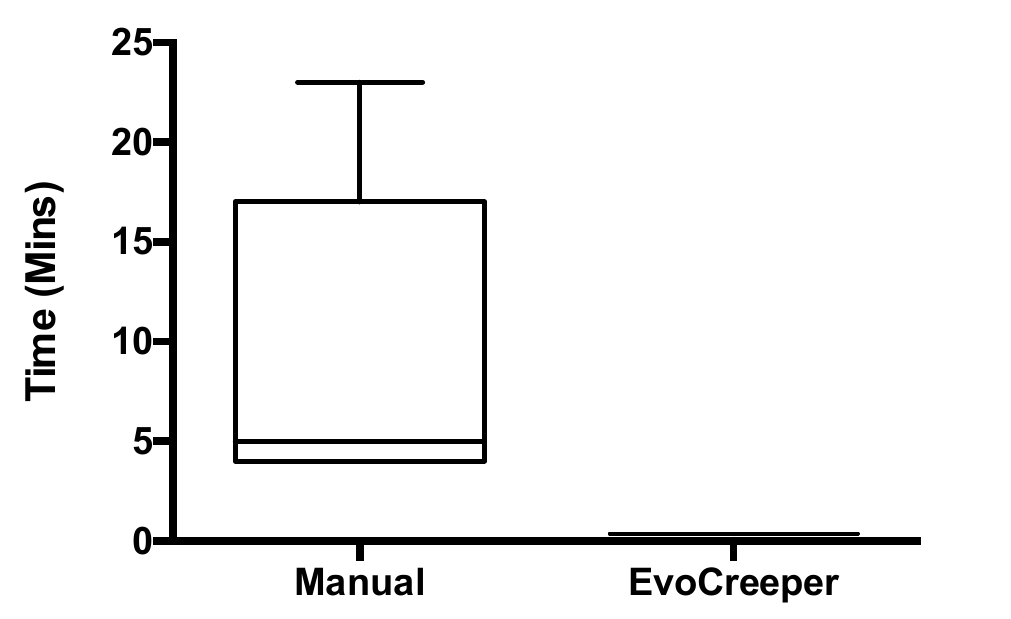}
                \vspace{-0.3cm}
		\caption{ChessLab \centering }
		\label{ChessLabTime}
    \end{subfigure}%
         \hfill
    \begin{subfigure}[b]{0.3\textwidth}
    	\includegraphics[width=\textwidth]					{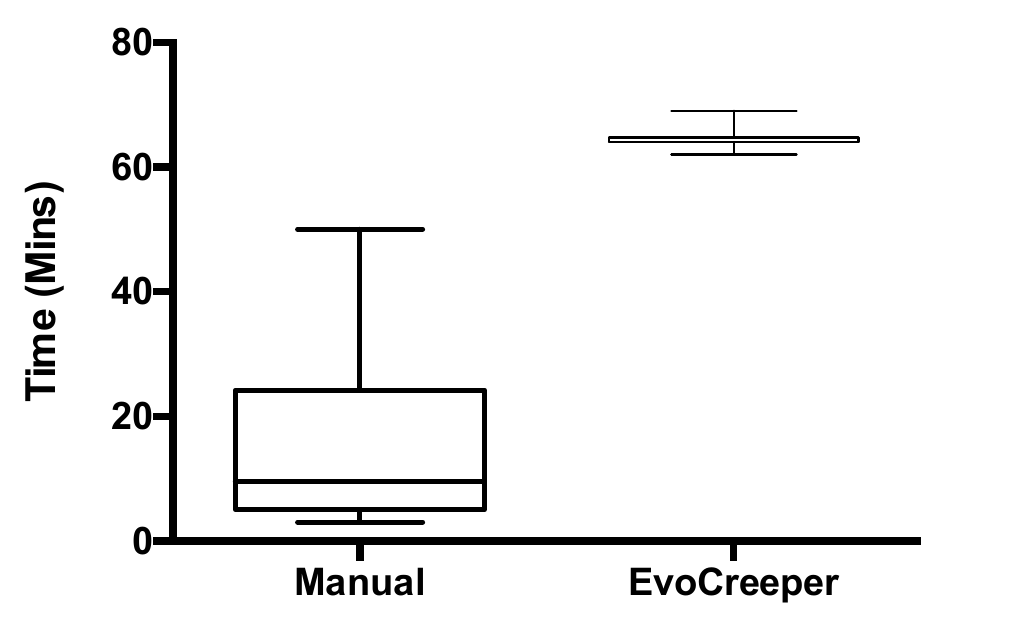}
                \vspace{-0.3cm}
        \caption{CineMup \centering }
        \label{CineMupTime}
          \end{subfigure}
             \hfill
             \begin{subfigure}[b]{0.3\textwidth}
    	\includegraphics[width=\textwidth]					{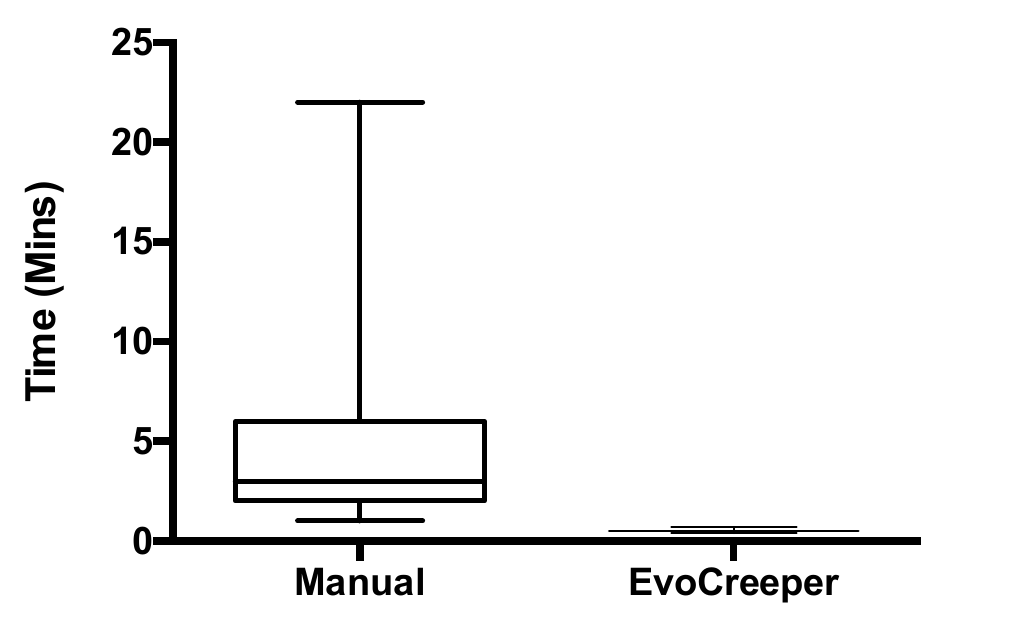}
                \vspace{-0.3cm}
        \caption{Memory game  \centering }
        \label{MemorygameTime}
          \end{subfigure}
             \hfill
\caption{Comparing the exploration time by EvoCreeper with the manual exploration}
\label{TimeNodeBoxPlot}
\end{figure*}

It is clear from the figures that our automated strategy can produce more accurate results by detecting more unique states (and as a result edges) on the GUI. It is also noticeable that our strategy can produce deterministic results as compared to manual exploration due to the different accuracy of the manual testers. This situation is clearer in the case of medium and large apps due to the many widgets on the GUI. From the experiments, we also observed the correctness of our strategy by verifying the generated models with the models generated manually by the testers. However, our strategy can generate more complete models by identifying more unique states and edges. We observed that for small-sized apps, the manual exploration also could be useful (depending on the tester experience), whereas, for the medium and large apps, the automated approach is necessary to obtain accurate models. For example, we can see in Fig. \ref{CineMupNodeNumber} that the best results achieved by the manual tester are worst than the automated approach. This situation is also evident in Fig. \ref{CineMupEdgeNumber} for the number of unique edges discovered. We observed during the experiments that the success of the manual exploration depends on the knowledge, experience, and accuracy of the tester. Hence, to escape from this prone-to-mistakes situation, it would be more practical to use the automated testing approach. While our strategy can achieve better results in terms of unique states and edges, it is also important to evaluate the model creation time and compare it with the manual model creation. Fig. \ref{TimeNodeBoxPlot} shows the box plot and whisker of this evaluation.

We observed from the experimental results that manual crawling time varied among the testers. This time variation is due to the knowledge and experience of the tester. Previous knowledge about the app also plays an important rule in the exploration time among the testers. The variation of the time from an app to another is due to the usability, level of complexity, and the size of the app. We noticed that the time variation range (it is clear from the whisker) is much larger than our automated crawler. For the medium-sized apps, the time required by the automated crawler is less than the median required time by the testers. This situation can be seen clearly in the case of the Memory app in Fig. \ref{MemorygameTime}. In other apps, the automated crawler takes less time for exploration. It should be mentioned here that the accuracy and the completeness must be taken into account also for the final evaluation conclusion. By revisiting the Fig. \ref{NumberOfUniqueNodeBoxPlot} and \ref{NumberOfUniqueEdgesBoxPlot} and combine the results with the results in Fig. \ref{TimeNodeBoxPlot}, it is clear that less exploration time does not necessary mean accuracy and completeness of the model. We observed that for large-sized apps, our strategy exploration is much deeper than the manual exploration due to the larger number of widgets; thus, it takes relatively more time. It can be seen clearly that the number of unique nodes and edges is much higher than the manual exploration of the app.


\subsection{Test Case Effectiveness and Mutation Testing}

In our testing framework, the model construction of the smart TV app is followed by a test generation phase. We have illustrated the test generation phase previously in this paper. To assess the effectiveness of the generated test cases, it is essential to see their usefulness term of fault detection. As mentioned earlier, the research in smart TV app testing is relatively new and we cannot find any mutation generation tools to inject faults into our case studies. We also cannot find any introduced mutation operators in the literature. To this end, in Section \ref{MutationOperators}, we have introduced new mutation operators for smart TV apps and that can be considered as an essential contribution in this paper. We have manually generated a list of those mutants and injected them into three case studies we used in this paper. In the three case study apps, we have injected 1753 mutants of those mutants presented in Table \ref{MutantTable}. It should be mentioned here that not all the mutants were applicable for all the apps. Table \ref{tab:Number-of-mutants} shows the number of mutants per each mutation operator used for each case study.

\begin{table}
\caption{Number of mutants per mutation operator used for each case study\label{tab:Number-of-mutants}}
\begin{centering}
\footnotesize
\begin{tabular}{|c|c|c|c|}
\hline 
Case Study & Mutation Oper. & No. & Total Number\tabularnewline
\hline 
\hline 
\multirow{4}{*}{CinemUp} & RAR & 89 & \multirow{4}{*}{1592}\tabularnewline
\cline{2-3} \cline{3-3} 
 & NEE & 49 & \tabularnewline
\cline{2-3} \cline{3-3} 
 & NEF & 1365 & \tabularnewline
\cline{2-3} \cline{3-3} 
 & NVR & 89 & \tabularnewline
\hline 
\multirow{4}{*}{CheesLab} & WRC & 28 & \multirow{4}{*}{146}\tabularnewline
\cline{2-3} \cline{3-3} 
 & BAV & 116 & \tabularnewline
\cline{2-3} \cline{3-3} 
 & IFC & 1 & \tabularnewline
\cline{2-3} \cline{3-3} 
 & NXE & 1 & \tabularnewline
\hline 
Memorry & NEA & 15 & 15\tabularnewline
\hline 
\multicolumn{3}{|l|}{Total} & 1753\tabularnewline
\hline 
\end{tabular}
\par\end{centering}
\end{table}

The testing process is undertaken using the Tizen SDK and its smart TV emulator. We have obtained the source code of the case studies to inject the mutants. We have generated the test cases from the sub-models. The test cases are sequences of steps to run a specific task on the smart TV app. We generated the test cases using Oxygen and directly converted them to scripts to run them on the SDK. 

In contrast to the mutation testing process for other apps, it is not possible to inject the mutants all together in the apps. Here, to detect a mutant during the execution of the test cases and the app, we mainly monitor the exception that must be thrown due to that mutant. This exception may stop the app in some particular cases (depending on the app under test) and the tester may need to restart the program. Hence, considered this situation for killing the mutants. For this reason, the testing had to be done in several phases so that the app passed all parts of the code according to the test path.

As mentioned previously, we have injected our mutation operators manually, not by using an automated mutation tool. This manual injection gives us the flexibility to inject the faults into the sub-model path not into the whole program randomly, as we are just testing the sub-model, not the mega-model. Hence, we choose a sub-model of the app; then we used that sub-model to generate the test cases. We then identify the path in the case study and inject the mutants into that path. For each generated test case (i.e., path), we recorded the killed mutants. As a result, we have calculated the total killed and alive mutants in addition to the mutation score. Table \ref{MutationScoreTable} shows the outcome of this empirical experiment for the three case studies.

\begin{table*}
\begin{centering}
\caption{Empirical results of the mutation testing\label{MutationScoreTable}}
\footnotesize
\begin{tabular}{|c|c|c|c|c|c|}
\hline 
Case Study & Generated Path & Mut. Killed & Total Mut. Killed & Mut. Alive & Total Mut. Score \%\tabularnewline
\hline 
\hline 
\multirow{3}{*}{CinemUp} & Path 1 & 655 & \multirow{3}{*}{1311} & \multirow{3}{*}{281} & \multirow{3}{*}{82.3}\tabularnewline
\cline{2-3} \cline{3-3} 
 & Path 2 & 414 &  &  & \tabularnewline
\cline{2-3} \cline{3-3} 
 & Path 3 & 242 &  &  & \tabularnewline
\hline 
\multirow{4}{*}{CheesLab} & Path 1 & 16 & \multirow{4}{*}{93} & \multirow{4}{*}{53} & \multirow{4}{*}{63.6}\tabularnewline
\cline{2-3} \cline{3-3} 
 & Path 2 & 15 &  &  & \tabularnewline
\cline{2-3} \cline{3-3} 
 & Path 3 & 46 &  &  & \tabularnewline
\cline{2-3} \cline{3-3} 
 & Path 4 & 16 &  &  & \tabularnewline
\hline 
Memorry & Path 1 & 9 & \multirow{2}{*}{12} & \multirow{2}{*}{3} & \multirow{2}{*}{80}\tabularnewline
\cline{1-3} \cline{2-3} \cline{3-3} 
 & Path 2 & 3 &  &  & \tabularnewline
\hline 
\end{tabular}
\par\end{centering}
\end{table*}

From Table \ref{MutationScoreTable}, we can see that for the CinemUp app, we have generated three paths based on the selected destination on the sub-model. While for the ChessLab there are four paths and for the Memorry app, there are two paths. The paths are sequences of states that must be executed automatically on the emulator using the remote control device. Here, we consider a mutant killed if it threw an exception in the log monitor. As can be seen clearly in the empirical results, each path can kill a certain number of mutants. We have recorded the number of those mutants killed by each path. It should be mentioned here that we excluded the repeated mutants in the killed number. We calculated the total killed mutants, alive mutants, and the mutation score. 

In the case of CinemUp app, the total killed number of mutants is 1311, while the number of alive mutants is 281, which leads to 82.3 \% mutation score. For the CheesLab app, the total killed mutants' number is 93 and the alive number is 53, which leads to a total of 63.6\% mutation score. Due to the small size of the Memorry app, we have only injected 15 mutants that we could detect 12 of them to lead to a total of 80\% mutation score. 

In total, we can see that the test cases created with the help of our model-based strategy were successful to detect faults in smart TV apps. It is also clear that our mutation strategy can be used as a base in the future to develop more mutation operators for smart TV apps and also to develop an automated mutation analysis tool for smart TV apps.

\section{Threats to Validity}\label{Threats}

There are a few threats that may affect the validity of the results in this paper. While we were aware of those threats, we tried to eliminate the effect of them when we conduct the experiments. 

Comparing the automated model generation algorithms' results with manual exploration is a possible threat to internal validity. As mentioned earlier, smart TV testing is an under-explored research area. As a result, we could not find a tool or a strategy that follows the same approach to re-engineering the app and generate the mega-model. Following the literature, an alternative and natural comparison solution are to compare with the manual exploratory testing approach \cite{REIS201849}. However, the manual exploration of the app could be subjective in some situations as it might be affected by the participant who is doing the testing. To eliminate this threat, we considered many participants per each experiment to assure fair results. We showed all those results achieved by the participants in terms of worst, best, and median in the box-blot graphs. The choice of the participants may affect the exploration time for example. This situation depends on the level of expertise of the participants, the usability of the explored app, and prior knowledge about the app. We have tried to eliminate this threat by choosing the best students in the computer science department and illustrating the workflow of the apps used for the case study. We have also given them enough time to explore the apps before the actual experiment starts. There could be other out of control human factors that may affect the exploration, model construction, and testing process. However, those factors give a strong reason to automate this testing process as a whole. If the tester does not have prior information about the app, he/she may produce the worst results that we have shown in this paper.

Running the experiments on the emulator, not a real smart TV device could also be a source of threats to validity. While the emulator represents an actual TV device, the exploration and the model generation of the smart TV app does not affect by the running environment. Hence, we expect to get the same model with both emulated and actual Smart TVs. Besides, the mutation testing process and the mutants are app-specific and they don't affect by the actual device. Testing the app with our approach on a real smart TV may lead to detect a different set of faults which may be related to the interaction of the app with the actual device. For example, a fault in the app may lead to restarting the smart TV device through its operating system. However, this kind of faults and testing approach is out of the scope of this paper.

Some of the smart TV apps may use more than those navigation buttons on the remote control device. This could be a threat that may lead to an incomplete model for a specific app. Those buttons are app-specific and the app developer may specify them. In most of the apps, those specific buttons are frequently used. In our model construction, we did not consider those individual cases as they are app specific and they may be added to our algorithm easily if we specifically know the key. This can be done by adding another navigation button. However, for the case studies used in this paper, we avoided using that kind of apps.

Another threat to validity may raise when the content of the smart TV app changes from the cloud that is lead to changing the GUI itself. This threat could form a critical situation if, for example, the experimental group run the empirical experiments in different days, that leads to changing the model of the app, which in term leads to different experimental results. To avoid this situation, we committed our self to run all the experiments on the same day both on our testing framework and with all the experimental group.

\section{Conclusion}\label{Conclusion}

In this paper, we have presented our automated testing framework to test smart TV apps. The framework relies on reverse engineering and crawling the smart TV apps to create a comprehensive model (mega-model) of it, considering the remote control device as the primary way of user interaction. The framework crawling the app constructing the model without knowing the internal source code of it, i.e., a black-box approach. Due to the enlargement of the model for medium and large size apps, we have also introduced the notion of model separation and sub-model. We have evaluated our approach using three case studies of small, medium, and large size smart TV apps. We have also assessed the effectiveness of the generated test cases on case studies through mutation testing.



\bibliographystyle{IEEEtran}
\bibliography{sample}

\end{document}